\documentclass[hyper]{jhep}
\newfont{\Bbb}{msbm10 scaled 1200}
\newcommand{\mbb}[1]{\mbox{\Bbb #1}} 
\title{Instantons and the monopole-like equations in eight dimensions}
\author{{Yi-hong Gao\thanks{permanent address: Institute of Theoretical
Physics, Beijing 100080, China} } and Gang Tian\\ Department of Mathematics,
Massachusetts Institute of Technology \\ 77 Massachusetts Avenue,
Cambridge, MA 02139 \\Email: \email{yhgao@math.mit.edu}, \email{tian@math.mit.edu}}
\abstract{We search for an abelian description of the Yang-Mills
instantons on certain eight dimensional manifolds with the special holonomies
$Spin(7)$ and $SU(4)$. By mimicing the Seiberg-Witten theory in four
dimensions, we propose a set of monopole-like equations governing the
8-dimensional $U(1)$ connections and spinors, which are supposed to be the
dual theory of the nonabelian instantons. We also give a naive test of
the generalized $S$-duality in the abelian sector of 8-dimensional
Yang-Mills theory. Some problems in this approach are pointed out.}

\preprint{hep-th/0004167} 
\begin{document}
\section{Introduction}
Yang-Mills instantons are among the simplest class of BPS states in the
low-energy limit of superstring theory. When strings are compactified, it is
often important to consider instantons on some special manifolds of dimension
other than four. Mathematically, such instantons arise naturally as solutions
to the eigenequations of a certain star operator acting on two-forms, and
just as in the 4-dimensional case, the Yang-Mills action will reach its
minimal values at these solutions. The present paper will be devoted to a
study of instantons in eight dimensions.

The notion of Yang-Mills instantons in dimension greater than four
is rather old and it may date back to the middle of the 1980's
\cite{CDFN}\cite{Ward}. This problem has raised some renewed interest in
recent years as a meanings of generalizing the Donaldson-Witten theory to
higher dimensions \cite{Oxford}. In particular, it is quite interesting to
see whether Donaldson invariants have the holomorphic extension to Calabi-Yau
four folds. Motivated by this as well as by the potential relevance to
M-theory and D-brane physics, various aspects of higher dimensional
cohomological Yang-Mills theories have been investigated, see
{\it e.g.} \cite{BKS}--\cite{AFSO}. An extensive study of the relevant
moduli geometry and its relations to certain calibrated submanifolds can be
found in ref.\cite{tian}. It is expected that the instanton configurations
should correspond to supersymmetric D-branes embedded in some manifolds of
special holonomies \cite{cycles}. In a more recent paper, Mari\~{n}o, Minasian,
Moore and Strominger \cite{MMMS} explicitly found that a nonlinear
deformation of the higher dimensional instanton equations can be derived from
D-branes wrapping around supersymetric cycles, with the deformation parameter
characterized by the $B$-field.

Presumably, the field theoretic approach to the instanton moduli problem
based on BRST cohomology \cite{BKS}\cite{AOS}\cite{AFSO} is perturbative
in nature. The quantum degrees of freedom consist mainly of the nonabelian
gauge fields $A$, which should be considered as fundamental fields when
we try to develop a perturbative expansion in terms of the gauge coupling
constant. One may ask whether there exists a nonperturbative theory
within which one can use collective field variables to explore the underlying
strong coupling physics \cite{BKS}. Inspired by the work of Seiberg and Witten \cite{SW}
in four dimensions, we tentatively expect that such a theory, if exists,
should be closely related to a kind of $S$-duality. Moreover, in the dual
description the collective variables should consist of an abelian gauge field
together with a complex spinor satisfying certain ``master equations''
\cite{W94}.

In this paper we take a modest step toward an $S$-dual description
for the Yang-Mills instantons on some eight dimensional manifolds with special
holonomy groups. Our description mimics the Seiberg-Witten theory \cite{W94},
in which the nonabelian (anti) self-duality equation $F^{+}=0$ will be
replaced by an abelian one, $F^{+}={\cal Q}(\psi^{\dag},\psi)$, with $\psi$
being a spinor field obeying the massless Dirac equation and $\cal Q$ a
suitable quadratic form. In writing down the explicit monopole-like equations,
we shall consider two types of manifolds: Joyce manifolds \cite{Joyce} of
holonomy $Spin(7)$ as well as Calabi-Yau four-folds of holonomy $SU(4)$. We
will compare our equations with the monopole equations constructed in
4-dimensions, and point out some problems yet to be resolved.

As a physical motivation of this investigation, we will also discuss the free
abelian sector embedded in 8-dimensional Yang-Mills theory and provide a
naive path-integral test of the $S$-duality in that sector. This discussion
is an eight dimensional generalization of the usual electric-magnetic duality
in four dimensions. The duality structure in eight dimensions may be alternatively
understood as the existence of different gauge-fixings of a topological symmetry
\cite{BS}.

The paper is organized as follows. In Section 2 we recall some known facts
about Yang-Mills instantons on manifolds of holonomy groups $Spin(7)$ and
$SU(4)$. In Section 3, we give an explicit construction of the monopole-like
equations. In Section 4 we turn to a discussion of the generalized $S$-duality
in 8-dimensional abelian gauge theory. Finally we provide an Appendix where
some useful properties of the 8-dimensional Clifford algebra are presented.
\section{Yang-Mills Instantons in Eight Dimensions}
Yang-Mills instantons in eight dimensions \cite{CDFN}\cite{Ward}\cite{BKS}
originate from a generalization of the usual concept of (anti) self-duality.
Suppose that we have an eight-dimensional Riemannian manifold $X$ on which a
closed 4-form $\Omega$ is defined. One can use this $\Omega$ to construct a star
operator $*_{\Omega}:\Lambda^{2}\rightarrow\Lambda^{2}$, $*_{\Omega}F\equiv
*(\Omega\wedge F)$ acting on the space of two-forms. The (anti) self-duality
equations are then formulated as the eigenequation $*_{\Omega}F=\lambda F$ of
the star operator.

In terms of components, the action of $*_{\Omega}$ is given by
\begin{equation}
(*_{\Omega}F)_{\mu\nu}=\frac{1}{2}\Omega_{\mu\nu\alpha\beta}F^{\alpha\beta}.
\label{star}
\end{equation}
Thus, if $\Omega$ obeys an identity of the form
\begin{equation}
\Omega_{\mu\nu\alpha\beta}\Omega^{\alpha\beta\sigma\tau}
=A(\delta_{\mu}^{\sigma}\delta_{\nu}^{\tau}-\delta_{\mu}^{\tau}
\delta_{\nu}^{\sigma})+B{\Omega_{\mu\nu}}^{\sigma\tau}
\label{OmegaID}
\end{equation}
with some real constants $A$ and $B$ (where $A>0$ depends on the normalization of
$\Omega$), then the eigenequation has two solutions $\lambda=\lambda^{\pm}$,
$F=F^{\pm}$, determined by:
\begin{equation}
\begin{array}{lll}
\lambda^{\pm} & = & \displaystyle\frac{B\mp\sqrt{B^{2}+8A}}{4},\vspace{.3cm}\\
F^{\pm}_{\mu\nu} & = & \displaystyle\pm\frac{2}{\sqrt{B^{2}+8A}}\left(\lambda^{\mp}
F_{\mu\nu}-\frac{1}{2}\Omega_{\mu\nu\alpha\beta}F^{\alpha\beta}\right).
\end{array}
\label{eigen}
\end{equation}
One may easily verify that $F_{\mu\nu}=F^{+}_{\mu\nu}+F^{-}_{\mu\nu}$.
Accordingly, the space $\Lambda^{2}$ of two-forms is decomposed into a direct
sum $\Lambda^{2}_{+}\oplus\Lambda^{2}_{-}$ of the eigenspaces of $*_{\Omega}$.
We will call $F$ to be a self-dual form (resp. anti self-dual form) if it
belongs to $\Lambda^{2}_{-}$ (resp. $\Lambda^{2}_{+}$). The condition that
$F$ is self-dual can be simply written as $F^{+}_{\mu\nu}=0$.

Actually we shall forcus on some $G$-bundle $E\rightarrow X$ and consider
its connections $A$. In this context, $A$ is called a self-dual instanton
(upto gauge transformations) if the corresponding curvature two-form $F(A)$
obeys the self-duality equation $F^{+}(A)=0$. An instanton will minimize the
Yang-Mills action functional
\begin{equation}
S_{YM}[A]=\frac{1}{2g^{2}}\int_{X} d^{8}x\sqrt{g}\,{\rm Tr}F_{\mu\nu}
F^{\mu\nu}\equiv \frac{1}{2g^{2}}||F||^{2}.
\label{YMaction}
\end{equation}
In fact, if we decompose $F$ into components $F^{\pm}\in\Lambda^{2}_{\pm}$,
then (\ref{eigen}) gives:
\begin{equation}
S_{YM}[A]=\frac{1}{g^{2}(B+\sqrt{B^{2}+8A})}\left(
2\int_{X}\Omega\wedge{\rm Tr}(F\wedge F)
+\sqrt{B^{2}+8A}||F^{+}||^{2}\right).
\end{equation}
So the Yang-Mills action can be written as a non-negative term
proportional to $||F^{+}||^{2}$, plus a topological invariant. Clearly, such
an action will reach its minimal values at $F^{+}=0$.
\vskip 1.2mm
\underbar{Manifolds with $Spin(7)$ holonomy}. Now we briefly discuss the
case when $X$ has the holonomy group $Spin(7)$. This means that $X$ is spin,
and there is a real, non-zero parallel spinor $\zeta\in S^{+}$ on $X$
invariant under the action of $Spin(7)\subset Spin(8)$. We will normalize
such a spinor by imposing the condition
\begin{equation}
\zeta^{T}\zeta=1.
\label{normalization}
\end{equation}

According to the standard isomorphism $S^{+}\otimes_{S} S^{+}\cong\Lambda^{0}
\oplus\Lambda^{4}_{+}$, $S^{+}\wedge S^{+}\cong\Lambda^{2}$ between the space
of forms and tensor product of the Clifford module \cite{spin}, the
4-form $\Omega$ considered above can be constructed as a ``bispinor''
\begin{equation}
\Omega_{\mu\nu\alpha\beta}=\zeta^{T}\Gamma_{\mu\nu\alpha\beta}\zeta,
\label{invTensor}
\end{equation}
where $\Gamma_{\mu\nu\alpha\beta}$ denotes the anti-symmetrized 4-fold product of
the $\gamma$-matrices $\Gamma_{\mu}$ in 8 dimensions, with a prefactor
$1/4!$ included. Our convention of choosing the $\gamma$-matrices is given in
the Appendix.

Eq.(\ref{invTensor}) obviously defines a $Spin(7)$ invariant rank-4 tensor.
This tensor enjoys a couple of useful properties: First, it is covariantly
constant, so that $\Omega$ gives rise to a closed form. Second, using the
$\gamma$-matrix identity $\Gamma_{\mu\nu\alpha\beta}\Gamma_{9}=\frac{1}{4!}
\epsilon_{\mu\nu\alpha\beta\lambda\rho\sigma\tau}\Gamma^{\lambda\rho\sigma
\tau}$, one easily sees that $\Omega$ is self-dual with respect to the usual
Hodge star operator, namely $*\Omega =\Omega$, -- this agrees with the fact that
the symmetric tensor product of $S^{+}$ contains $\Lambda^{4}_{+}$. The final
property which we shall use is that (\ref{invTensor}) obeys an identity
\cite{AFSO} of the form (\ref{OmegaID}):
\begin{equation}
\Omega_{\mu\nu\alpha\beta}\Omega^{\alpha\beta\sigma\tau}
=6(\delta_{\mu}^{\sigma}\delta_{\nu}^{\tau}-\delta_{\mu}^{\tau}
\delta_{\nu}^{\sigma}) -4{\Omega_{\mu\nu}}^{\sigma\tau}.
\label{OmegaID-R}
\end{equation}
In particular $\Omega$ is normalized to be $||\Omega||^{2}\equiv\frac{1}
{4!}\Omega_{\mu\nu\alpha\beta}\Omega^{\mu\nu\alpha\beta}=14$.

Hence, on a manifold $X$ of holonomy $Spin(7)$, the (anti) self-duality
equation $*_{\Omega}F=\lambda F$ has eigenvalues $\lambda^{+}=-3$,
$\lambda^{-}=1$; the curvature two-form $F_{\mu\nu}=F_{\mu\nu}^{+}+
F_{\mu\nu}^{-}$ is decomposed orthogonally into an anti self-dual part
$F^{+}_{\mu\nu}$ and a self-dual part $F^{-}_{\mu\nu}$, with
\begin{equation}
\begin{array}{lll}
F^{+}_{\mu\nu} & = & \displaystyle\frac{1}{4}\left(F_{\mu\nu}-\frac{1}{2}
\Omega_{\mu\nu\alpha\beta}F^{\alpha\beta}\right)\in\Lambda^{2}_{+},
\vspace{.3cm}\\
F^{-}_{\mu\nu} & = & \displaystyle\frac{1}{4}\left(
3F_{\mu\nu}+\frac{1}{2}\Omega_{\mu\nu\alpha\beta}F^{\alpha\beta}\right)
\in\Lambda^{2}_{-}.
\end{array}
\label{eigen-R}
\end{equation}
The Yang-Mills instantons are thus described by the equation $F^{+}_{\mu\nu}
=0$. The dimensions of $\Lambda^{2}_{\pm}$ can be determined by a
group-theoretic
consideration \cite{BKS}, and the result turns out to be
\begin{equation}
\dim\Lambda^{2}_{+}=7,\quad\dim\Lambda^{2}_{-}=21.
\label{dim}
\end{equation}
For an alternative derivation of this result, note that ${\rm Tr}(*_{\Omega})
=0$, so $(-3)\cdot\dim\Lambda^{2}_{+}+1\cdot\dim\Lambda^{2}_{-}=0$. This
together with $\dim\Lambda^{2}_{+}+\dim\Lambda^{2}_{-}=\dim\Lambda^{2}=28$
gives (\ref{dim}).
\vskip 1.2mm
\underbar{Manifolds with $SU(4)$ holonomy}. As just mentioned, $X$ has 
holonomy $Spin(7)$ if there is a generic parallel spinor $\zeta\neq 0$
defined on it. However, this holonomy group may reduce to a subgroup of
$Spin(7)$ when the parallel spinor obeys certain particular conditions
\cite{spin}. For example, one has the holonomy reduction $Spin(7)
\rightarrow SU(4)$ provided there exists a parallel {\it pure spinor} $\zeta$
on $X$. Here we shall describe in some detail what a pure spinor is and
explain why the existence of such a spinor will cause the manifold to have
holonomy $SU(4)$ \cite{spin}. We will then discuss a holomorphic version of
the Yang-Mills instanton equations \cite{BKS}.

Let $Cl_{c}(8)=Cl(8)\otimes\mbb C$ be the 8-dimensional Clifford algebra over
$\mbb C$, and let $S_{c}$ be a complex spinor space, on which an irreducible
representation $\rho^{c}$ of $Cl_{c}(8)$ is defined. Each spinor $\psi\in
S_{c}$ can be associated to a $\mbb C$-linear map
\begin{equation}
f_{\psi}: \mbb C^{8}\rightarrow S_{c},\quad f_{\psi}(u)=\rho^{c}(u)\cdot\psi,
\label{c-map}
\end{equation}
where $u$ is a complex linear combination of the Clifford generators
$e_{\mu}\in Cl(8)$ and we have identified the space of all such linear
combinations with $\mbb C^{8}$. Let us consider the kernel of this map
in the case $\psi\neq 0$. If $u=a^{\mu}e_{\mu}$, $v=b^{\mu}e_{\mu}\in
{\rm Ker}\,f_{\psi}$ (with complex coefficients $a^{\mu}$, $b^{\mu}$),
then $\rho^{c}(u)\psi=\rho^{c}(v)\psi=0\Rightarrow 0=\rho^{c}(\{u,v\})\psi=
-2g_{\mu\nu}a^{\mu}b^{\nu}\psi$. It follows that the space ${\rm Ker}\,f_{\psi}$
is orthogonal to its complex conjugate $\overline{{\rm Ker}\,f_{\psi}}$ with
respect to the standard hermitian inner product $(a^{\mu}e_{\mu},b^{\nu}
e_{\nu})\equiv\langle a^{\mu}e_{\mu},\bar{b}^{\nu}e_{\nu}
\rangle= g_{\mu\nu}a^{\mu}\bar{b}^{\nu}$ on $\mbb C^{8}$. Thus, since
${\rm Ker}\,f_{\psi}\oplus\overline{{\rm Ker}\,f_{\psi}}\subset\mbb C^{8}$,
we see that the complex dimensions of ${\rm Ker}\,f_{\psi}$ should not
exceed 4. By definition, we say that $\psi$ is a pure spinor if
$\dim_{c}{\rm Ker}\,f_{\psi}$ reaches its maximally allowed value 4.

We will take the complex spinor space to be the complexification of
the real $Cl(8)$-module $S\cong\mbb R^{16}$: $S_{c}=S\otimes\mbb C$. Spinors
in such a space can be written as linear combinations of a basis of $S$
with complex coefficients. Also, one takes $\rho^{c}$ to be the $\mbb
C$-linear extension of the $\gamma$-matrix representation of $Cl(8)$. This
allows us to choose $\Gamma_{\mu}\equiv\rho^{c}(e_{\mu})$ as given in the
Appendix.

Since $\Gamma_{9}=I\oplus(-I)$, $S_{c}$ is decomposed into subspaces
$S_{c}^{\pm}$ of positive and negative chiralities. By
construction, $\psi\in S_{c}^{+}$ means that $\psi$ is a linear combination
of some real spinors in $S^{+}$ with complex coefficients, so its complex
conjugate $\bar{\psi}$ also has positive chirality. One can show
that \cite{spin} if $\psi$ is a pure spinor, then either $\psi$ will be
entirely in $S_{c}^{+}$ or it will be entirely in $S_{c}^{-}$; namely, $\psi$ has a
definite chirality. To see this, note that a change in the orthonormal basis
$\{e_{\mu}\}$ of $\mbb R^{8}\subset Cl(8)$ will leave the matrix $\Gamma_{9}=
\rho^{c}(e_{1}\cdots e_{8})$ invariant, upto a factor $\pm 1$ depending on
the relative ordering. Also note that if $\psi$ is a pure spinor, then
${\rm Ker}\,f_{\psi}\oplus\overline{{\rm Ker}\,f_{\psi}}=\mbb C^{8}$, so any
basis $\{h_{\bar{i}},1\leq\bar{i}\leq 4\}$ of ${\rm Ker}\,f_{\psi}$ along
with its complex conjugate $\{h_{i}\}\subset\overline{{\rm Ker}\,f_{\psi}}$
provides a basis of $\mbb C^{8}$. We can take $\{h_{\bar{i}}\}$ to be
orthonormal with respect to the natural hermitian metric on $\mbb C^{8}$.
Then the following vectors
\begin{equation}
\begin{array}{llll}
\displaystyle{e_{1}=\frac{1}{\sqrt{2}}(h_{1}+h_{\bar{1}}),}&
\displaystyle{e_{2}=\frac{1}{\sqrt{2}i}(h_{1}-h_{\bar{1}}),}&
\displaystyle{e_{3}=\frac{1}{\sqrt{2}}(h_{2}+h_{\bar{2}}),}&
\displaystyle{e_{4}=\frac{1}{\sqrt{2}i}(h_{2}-h_{\bar{2}})}
\vspace{.3cm}\\
\displaystyle{e_{5}=\frac{1}{\sqrt{2}}(h_{3}+h_{\bar{3}}),}&
\displaystyle{e_{6}=\frac{1}{\sqrt{2}i}(h_{3}-h_{\bar{3}}),}&
\displaystyle{e_{7}=\frac{1}{\sqrt{2}}(h_{4}+h_{\bar{4}}),}&
\displaystyle{e_{8}=\frac{1}{\sqrt{2}i}(h_{4}-h_{\bar{4}})}
\end{array}
\label{basis}
\end{equation}
form an orthonormal basis of $\mbb R^{8}$, as they are all invariant under
complex conjugation. Now as $h_{\bar{1}}\in{\rm Ker}\,f_{\psi}$, we have
$\rho^{c}(e_{1}-ie_{2})\psi=0\Rightarrow\rho^{c}(e_{1}^{2}-ie_{1}e_{2})\psi=0
\Rightarrow\rho^{c}(e_{1}e_{2})\psi=i\psi$. Similar arguments lead to
$\rho^{c}(e_{3}e_{4})\psi=\rho^{c}(e_{5}e_{6})\psi=\rho^{c}(e_{7}e_{8})\psi=
i\psi$. Thus we find $\Gamma_{9}\psi=\psi$, indicating that $\psi$ has the
positive chirality. A choice of the basis with different ordering will give
$\Gamma_{9}\psi=-\psi$, but in any case the chirality of a pure spinor is
definite.

Given now a pure spinor $\zeta\in S^{+}_{c}$, let us consider the maximal
subgroup of $Spin(8)$ that keeps $\zeta$ invariant. An element of $Spin(8)$
can act adjointly on the real vector space spanned by $\{e_{\mu}\}$,
\begin{equation}
e_{\mu}\rightarrow e'_{\mu}=g^{-1}e_{\mu}g\equiv{\rho_{v}(g)_{\mu}}^{
\nu}e_{\nu},\quad \rho_{v}(g)\in SO(8),
\label{AdjAct}
\end{equation}
which defines the representation ${\bf 8}_{v}$ of $Spin(8)$. The action
(\ref{AdjAct}) may be viewed as a change in the basis and it clearly
preserves both the orthonormal property and the ordering of the basis. The
$\mbb C$-linear extension of $\rho_{v}$, which we will denote by
$\rho^{c}_{v}$, is defined naturally on the space $\mbb C^{8}={\rm Ker}\,
f_{\zeta}\oplus\overline{{\rm Ker}\,f_{\zeta}}$. For generic $g\in Spin(8)$,
neither the subspace ${\rm Ker}\,f_{\zeta}$ nor $\overline{{\rm Ker}\,
f_{\zeta}}$ is invariant under the action of $\rho^{c}_{v}(g)$.
Elements of $SO(8)$ that leaves these subspaces invariant
will map one orthonormal basis $\{h_{i}\}\subset\overline{{\rm Ker}\,
f_{\zeta}}$ (and $\{h_{\bar{i}}\}\subset {\rm Ker}\,f_{\zeta}$) into another,
thus forming the subgroup $SU(4)$. Such elements arise from those $g\in
Spin(8)$ keeping $\zeta$ invariant. Indeed, for any $h\in {\rm Ker}\,
f_{\zeta}$ and $\rho^{c}(g)\zeta=\zeta$, we have $\rho^{c}(g^{-1}hg)\zeta=
\rho^{c}(g)^{-1}\rho^{c}(h)\rho^{c}(g)\zeta=\rho^{c}(g)^{-1}\rho^{c}(h)\zeta
=0\Rightarrow g^{-1}hg\in{\rm Ker}\,f_{\zeta}$. We thus conclude that the
isotropy group of a pure spinor $\zeta\in S^{+}_{c}$ is $SU(4)$. A
globalized version of this discussion leads to the statement \cite{spin}:

{\it There exists a parallel pure spinor on $X\Longleftrightarrow$ $X$ has the
holonomy group $SU(4)$ (or its subgroup)}.

Now we take a pure spinor $\zeta\in S^{+}_{c}$ and fix the almost complex
structure on $\mbb R^{8}$ as in (\ref{basis}), so that the basis $h_{i}$ of
$\overline{{\rm Ker}\,f_{\zeta}}$ and the basis $h_{\bar{i}}$ of
${\rm Ker}\,f_{\zeta}$ have the $\gamma$-matrix representation:
\begin{equation}
\gamma_{1}=\frac{1}{\sqrt{2}}(\Gamma_{1}+i\Gamma_{2}),
\gamma_{2}=\frac{1}{\sqrt{2}}(\Gamma_{3}+i\Gamma_{3}),
\gamma_{3}=\frac{1}{\sqrt{2}}(\Gamma_{5}+i\Gamma_{6}),
\gamma_{4}=\frac{1}{\sqrt{2}}(\Gamma_{7}+i\Gamma_{8})
\label{gamma+}
\end{equation}
\begin{equation}
\gamma_{\bar{1}}=\frac{1}{\sqrt{2}}(\Gamma_{1}-i\Gamma_{2}),
\gamma_{\bar{2}}=\frac{1}{\sqrt{2}}(\Gamma_{3}-i\Gamma_{3}),
\gamma_{\bar{3}}=\frac{1}{\sqrt{2}}(\Gamma_{5}-i\Gamma_{6}),
\gamma_{\bar{4}}=\frac{1}{\sqrt{2}}(\Gamma_{7}-i\Gamma_{8}).
\label{gamma-}
\end{equation}
The complex Clifford algebra is determined by the relations
\begin{equation}
\gamma_{i}\gamma_{j}+\gamma_{j}\gamma_{i}=\gamma_{\bar{i}}\gamma_{\bar{j}}
+\gamma_{\bar{j}}\gamma_{\bar{i}}=0,\quad
\gamma_{i}\gamma_{\bar{j}}+\gamma_{\bar{j}}\gamma_{i}=-2g_{i\bar{j}}.
\label{Cliif-C}
\end{equation}
We also need the dual basis
\begin{equation}
\gamma^{i}=g^{i\bar{j}}\gamma_{\bar{j}},\quad
\gamma^{\bar{i}}=\gamma_{j}g^{j\bar{i}}
\label{dualBasis}
\end{equation}
as well as their anti-symmetrized products $\gamma^{i_{1}\cdots i_{p}\bar{j}_{1}
\cdots\bar{j}_{q}}$. With this notation, a $(p,q)$-form $t\in\Lambda^{p,q}$
has a natural representation in terms of the $\gamma$-matrices:
\begin{equation}
t\leftrightarrow
\frac{1}{p!q!}t_{i_{1}\cdots i_{p}\bar{j}_{1}\cdots\bar{j}_{q}}
\gamma^{i_{1}\cdots i_{p} \bar{j}_{1}\cdots\bar{j}_{q}}.
\label{pqForm}
\end{equation}
Moreover, each such form should be associated to a bispinor $\phi^{\dag}
\gamma_{i_{1}\cdots i_{p}\bar{j}_{1}\cdots\bar{j}_{q}}\psi\in S_{c}\otimes
S_{c}$ as in the real case. Note that the isomorphism \cite{spin} between the
tensor product of spinors and forms has a $\mbb C$-bilinear extension to
the complex case
\begin{equation}
\rho^{c}\otimes \rho^{c}\cong 2(1+\rho^{c}_{v}+\wedge^{2}\rho^{c}_{v}
+\wedge^{3}\rho^{c}_{v})+\wedge^{4}\rho^{c}_{v},
\label{isomComplex}
\end{equation}
which shows that $(S^{+}_{c}\oplus S^{-}_{c})\otimes
(S^{+}_{c}\oplus S^{-}_{c})$ can be identified with forms in $\wedge^{*}
\mbb C^{8}$.

To warm up the complex Clifford calculus, let us establish an isomorphism
between $S_{c}^{\pm}\otimes\zeta^{\dag}$ and certain particular forms. Since
$\gamma_{\bar{i}}$ is in ${\rm Ker}f_{\zeta}$, we have
\begin{equation}
\gamma_{\bar{i}}\zeta=\gamma^{i}\zeta=0\quad\Rightarrow\quad
\zeta^{\dag}\gamma_{i}=\zeta^{\dag}\gamma^{\bar{i}}=0,
\label{annihilation}
\end{equation}
and this gives to $\zeta^{\dag}\gamma_{i}\gamma_{i_{1}\cdots i_{p}\bar{j}_{1}
\cdots\bar{j}_{q}}=0$. One may use this and the $\gamma$-matrix identity
\begin{equation}
\gamma_{i}\gamma_{i_{1}\cdots i_{p}\bar{j}_{1}\cdots\bar{j}_{q}}=
\gamma_{ii_{1}\cdots i_{p}\bar{j}_{1}\cdots\bar{j}_{q}}+
\sum_{k=1}^{q}(-1)^{k+p}g_{i\bar{j}_{k}}\gamma_{i_{1}\cdots i_{p}
\bar{j}_{1}\cdots\widehat{\bar{j}_{k}}\cdots\bar{j}_{q}}
\label{gammaID-C}
\end{equation}
to deduce that the $(p+1,q)$ type bispinor $\zeta^{\dag}\gamma_{ii_{1}\cdots
i_{p}\bar{j}_{1}\cdots\bar{j}_{q}}\psi$ is in fact a linear combination of
some $(p,q-1)$ forms. This process can be proceeded inductively and we find
that the $(p+1,q)$-bispinor finally becomes a linear combination
of $\zeta^{\dag}\gamma_{\bar{j}_{1}\cdots\bar{j}_{q-p-1}}\psi$ if $q\geq p+1$
or a linear combination of $\zeta^{\dag}\gamma_{i_{1}\cdots i_{p+1-q}}\psi
\equiv 0$ if $q<p+1$. Accordingly, for arbitrary $\psi\in S_{c}=S^{+}_{c}
\oplus S^{-}_{c}$, the tensor product $\zeta^{\dag}\otimes\psi$ can be
identified to a form in $\Lambda^{0,*}$. Note that with our convention of
the $\gamma$-matrices, $\gamma_{\bar{j}_{1}\cdots\bar{j}_{q}}$ is
block diagonal for $q=$ even and off-diagonal for $q=$ odd. It follows that
\begin{equation}
S^{+}_{c}\otimes\mbb C\cong\Lambda^{0,{\rm even}},\quad
S^{-}_{c}\otimes\mbb C\cong\Lambda^{0,{\rm odd}}
\label{isomComplex1}
\end{equation}
here $\mbb C$ is the complex 1-dimensional space generated by $\zeta^{\dag}$.
Similarly, tensor products $\psi_{\pm}\otimes\zeta$ for $\psi_{\pm}\in S^{\pm
}_{c}$ should be identified with a form in $\Lambda^{{\rm even},0}$ and in
$\Lambda^{{\rm odd},0}$, respectively.

Now we give a suitable normalization of $\zeta$. In the $Spin(7)$ case we
have simply imposed the condition $\zeta^{T}\zeta=1$. However, this
normalization condition cannot be adopted here for a pure spinor $\zeta$. In
fact from (\ref{annihilation})-(\ref{gammaID-C}) we see that $\gamma_{i\bar{j}}
\zeta=g_{i\bar{j}}\zeta$, so that $\zeta^{T}\gamma_{i\bar{j}}\zeta=
g_{i\bar{j}}(\zeta^{T}\zeta)$, which together with the anti-symmetric
property of the matrix $\gamma_{i\bar{j}}$ implies $\zeta^{T}\zeta=0$.
Nevertheless, one can still impose another normalization condition
\begin{equation}
\zeta^{\dag}\zeta=1,
\label{|z|^2}
\end{equation}
and this looks more natural when we work in complex spaces. Using
this normlization, we define an $SU(4)$ invariant closed (4,0)-form $\Omega$
with the components
\begin{equation}
\Omega_{ijkl}=\zeta^{T}\gamma_{ijkl}\zeta.
\label{invTensorC}
\end{equation}

Some properties of (\ref{invTensorC}) can be explored using a
complex version of the Fierz rearrangement formula:
\begin{equation}
\zeta\zeta^{T}=\frac{1}{16\cdot 4!}\Omega_{ijkl}\gamma^{ijkl},\quad\quad\quad
\quad\quad\quad\quad\quad\quad\;
\label{fierz3-1}
\end{equation}
\begin{equation}
\zeta\zeta^{\dag} =
\frac{1}{16}(1+g^{i\bar{j}}\gamma_{i\bar{j}})(1+
\Gamma_{9})+\frac{1}{32}g^{i\bar{l}}g^{j\bar{k}}\gamma_{ij\bar{k}\bar{l}}.
\label{fierz3-2}
\end{equation}
For example one may apply (\ref{fierz3-1}) to a quick computation of the
norm $||\Omega||^{2}$. By definition, $||\Omega||^{2}\equiv\frac{1}{4!}
\Omega_{ijkl}\bar{\Omega}^{ijkl}$, $\bar{\Omega}^{ijkl}\equiv g^{i\bar{i}}
g^{j\bar{j}}g^{k\bar{k}}g^{l\bar{l}}\bar{\Omega}_{\bar{i}\bar{j}\bar{k}
\bar{l}}$, where $\bar{\Omega}_{\bar{i}\bar{j}\bar{k}\bar{l}}\in\Lambda^{0,4}$
is the complex conjugate of $\Omega$. Since $\bar{\Omega}^{ijkl}=\zeta^{
\dag}\gamma^{ijkl}\bar{\zeta}$, we see that (\ref{fierz3-1}) gives
$||\Omega||^{2}=16$. Notice that this normalization of $\Omega$ is different
from the $Spin(7)$ case, where $||\Omega||^{2}=14$. As an application of
(\ref{fierz3-2}), one can establish a more useful identity
\begin{equation}
\Omega_{ijkl}\bar{\Omega}^{mnkl}=32(\delta^{m}_{i}\delta^{n}_{j}
-\delta^{n}_{i}\delta^{m}_{j}),
\label{OmegaID-C}
\end{equation}
which takes a form similar to (\ref{OmegaID}).

We turn now to the self-duality equations. Given the $SU(4)$ invariant
(4,0)-form $\Omega_{ijkl}$ defined as above, its complex conjugate
$\bar{\Omega}_{\bar{i}\bar{j}\bar{k}\bar{l}}$, a (0,4)-form, may be used
to construct an anti-linear star operator $*_{\Omega}:\Lambda^{0,2}
\rightarrow\Lambda^{0,2}$ by means of
\begin{equation}
(*_{\Omega}\beta)_{\bar{i}\bar{j}}=\frac{1}{2}\bar{\Omega}_{\bar{i}\bar{j}
\bar{k}\bar{l}}\bar{\beta}^{\bar{k}\bar{l}},\quad\forall\beta_{\bar{i}\bar{j}}
\in\Lambda^{0,2},
\label{Ostar}
\end{equation}
where $\bar{\beta}_{ij}\in\Lambda^{2,0}$ denotes the complex conjugate of
$\beta_{\bar{i}\bar{j}}$ and $\bar{\beta}^{\bar{k}\bar{l}}\equiv g^{i\bar{k}}
g^{j\bar{l}}\bar{\beta}_{ij}$. Thus, if $F\in\Lambda^{2}$ is a curvature
2-form, we can decompose it into $F=F^{(2,0)}+F^{(1,1)}+F^{(0,2)}$ with
$F^{(2,0)}=-\overline{F^{(0,2)}}$ (assuming that the connection is unitary),
and define
\begin{equation}
(*_{\Omega}F^{(0,2)})_{\bar{i}\bar{j}}=-\frac{1}{2}\bar{\Omega}_{\bar{i}\bar{j}
\bar{k}\bar{l}}F^{(2,0)\bar{k}\bar{l}},\quad
(*_{\Omega}F^{(2,0)})_{ij}=-\frac{1}{2}\Omega_{ijkl}F^{(0,2)kl}.
\label{Ostar'}
\end{equation}
Note that the (1,1)-component of $F$ is intact under the action of
$*_{\Omega}$.

Just as in the $Spin(7)$ case, the (anti) self-duality equations should be
formulated as the eigenvalue equation of $*_{\Omega}$. Hence, in terms of
components, we call $F_{\mu\nu}$ to be (anti) self-dual if they satisfy the
conditions
\begin{equation}
-\frac{1}{2}\bar{\Omega}_{\bar{i}\bar{j}\bar{k}\bar{l}}F^{(2,0)\bar{k}\bar{l}}
=\lambda F^{(0,2)}_{\bar{i}\bar{j}},\quad
-\frac{1}{2}\Omega_{ijkl}F^{(0,2)kl}=\lambda F^{(2,0)}_{ij}.
\label{selfdualC}
\end{equation}
Here the eigenvalues $\lambda\in\mbb R$ are determined by
\begin{equation}
\lambda=\lambda_{\pm},\quad\quad \lambda_{+}=-4,\quad\lambda_{-}=4.
\label{eigenvalues}
\end{equation}
Accordingly, the space of (0,2)-forms gets decomposed into the two eigenspaces
of $*_{\Omega}$, $\Lambda^{0,2}=\Lambda^{0,2}_{+}\oplus\Lambda^{0,2}_{-}
$, where $\Lambda^{0,2}_{\pm}$ correspond  to the eigenvalues
$\lambda_{\pm}$, respectively. The (0,2) component of $F\in\Lambda^{2}$ then
decomposes into an anti self-dual part $F^{(0,2)}_{+}\in\Lambda_{+}^{0,2}$
and a self-dual part $F^{(0,2)}_{-}\in\Lambda_{-}^{0,2}$, with
\begin{equation}
F^{(0,2)}_{\pm\bar{i}\bar{j}}=\frac{1}{2}\left(F^{(0,2)}_{\bar{i}\bar{j}}
\pm\frac{1}{8}\bar{\Omega}_{\bar{i}\bar{j}\bar{k}\bar{l}}F^{(2,0)\bar{k}
\bar{l}}\right).
\label{decom+-}
\end{equation}
Holomorphic Yang-Mills instantons are thus characterized by the
self-duality equation
\begin{equation}
F^{(0,2)}_{+}(A)=0\quad\Longleftrightarrow\quad
-\frac{1}{2}\bar{\Omega}_{\bar{i}\bar{j}\bar{k}\bar{l}}F^{(2,0)\bar{k}\bar{l}}
=4F^{(0,2)}_{\bar{i}\bar{j}}.
\label{selfdualC-1}
\end{equation}                                                           

Sometimes it is useful to have a more compact description for the $*_{\Omega}$
operator, without reference to the unitary basis given in (\ref{basis}). To
give such a description, note that there is a natural hermitian inner product
on $\Lambda^{0,q}$: for two arbitrary $(0,q)$ forms $\alpha_{\bar{i}_{1}\cdots
\bar{i}_{q}}$ and $\beta_{\bar{i}_{1}\cdots\bar{i}_{q}}$, we can define an
$SU(4)$ invariant paring
\begin{equation}
\langle\alpha,\beta\rangle\equiv\frac{1}{q!}\alpha_{\bar{i}_{1}\cdots
\bar{i}_{q}}\bar{\beta}^{\bar{i}_{1}\cdots\bar{i}_{q}}=\frac{1}{q!}
g^{i_{1}\bar{j}_{1}}\cdots g^{i_{q}\bar{j}_{q}}\bar{\beta}_{i_{1}\cdots
i_{q}}\alpha_{\bar{j}_{1}\cdots\bar{j}_{q}},
\label{inner}
\end{equation}
which is linear in $\alpha$ and anti-linear in $\beta$. In terms of this
inner product, one then introduces an operator $*_{\Omega}:\Lambda^{0,q}
\rightarrow\Lambda^{0,4-q}$ through \cite{BKS}
\begin{equation}
\alpha\wedge *_{\Omega}\beta = \langle\alpha,\beta\rangle\bar{\Omega}.
\label{starC}
\end{equation}
Clearly, this description manifests the $SU(4)$ invariance and does not
depend on a particular choice of the basis of ${\rm Ker}f_{\zeta}$.
One may see that this definition agrees with the previous one for $q=2$ .

Actually, it is possible to consider a slightly generalized case where we
have an $SU(n)$ invariant $(n,0)$-form $\Omega_{i_{1}\cdots i_{n}}$
defined on some $2n$-dimensional space $X$.  In that case, the star operator
constructed by (\ref{starC}) should map $\beta\in\Lambda^{0,q}$ into
$*_{\Omega}\beta\in\Lambda^{0,n-q}$, so that $\alpha\wedge *_{\Omega}\beta$
is a $(0,n)$-form. The component of the left hand side of (\ref{starC}) is
$\frac{1}{q!(n-q)!}\alpha_{[\bar{i}_{1}\cdots\bar{i}_{q}}(*_{\Omega}\beta)_{
\bar{i}_{q+1}\cdots\bar{i}_{n}]}$, while the component of the right hand side
of (\ref{starC}) is $\frac{1}{n!q!}\alpha_{\bar{j}_{1}\cdots\bar{j}_{q}}\bar{
\beta}^{\bar{j}_{1}\cdots\bar{j}_{q}}\bar{\Omega}_{\bar{i}_{1}\cdots\bar{i}_{
n}}$; making them equal to each other for arbitrary $\alpha\in\Lambda^{0,q}$
leads to
$$
\delta^{\bar{j}_{1}}_{[\bar{i}_{1}}\cdots\delta^{\bar{j}_{n}}_{\bar{i}_{n}]}
(*_{\Omega}\beta)_{\bar{j}_{q+1}\cdots\bar{j}_{n}}
=\frac{(n-q)!}{n!}\bar{\beta}^{\bar{j}_{1}\cdots\bar{j}_{q}}
\bar{\Omega}_{\bar{i}_{1}\cdots\bar{i}_{n}}.
$$
By contracting the $q$ pairs $(\bar{i}_{1},\bar{j}_{1}),\cdots,(\bar{i}_{q},
\bar{j}_{q})$ of the tensor indices in this equation, and then using the
identity
$$
\delta^{\bar{i}_{1}}_{[\bar{i}_{1}}\cdots\delta^{\bar{i}_{q}}_{\bar{i}_{q}}
\delta^{\bar{j}_{q+1}}_{\bar{i}_{q+1}}\cdots\delta^{\bar{j}_{n}}_{
\bar{i}_{n}]}=\frac{q!(n-q)!}{n!}\delta^{\bar{j}_{q+1}}_{[\bar{i}_{q+1}}
\cdots\delta^{\bar{j}_{n}}_{\bar{i}_{n}]},
$$
we see that
$$
(*_{\Omega}\beta)_{\bar{i}_{q+1}\cdots\bar{i}_{n}}=\frac{1}{q!}\bar{\beta}^{
\bar{i}_{1}\cdots\bar{i}_{q}}\bar{\Omega}_{\bar{i}_{1}\cdots\bar{i}_{q}\bar{i
}_{q+1}\cdots\bar{i}_{n}}=\frac{(-1)^{q(n-q)}}{q!}\bar{\Omega}_{\bar{i}_{q+1}
\cdots\bar{i}_{n}\bar{i}_{1}\cdots\bar{i}_{q}}\bar{\beta}^{\bar{i}_{1}\cdots
\bar{i}_{q}}.
$$
In particular for $n=4$ and $q=2$, this reduces to our earlier definition
(\ref{Ostar}).
\section{The Monopole-like Equations}
Non-abelian instantons constitute a moduli problem. In 4-dimensions, this
problem can be transformed into a simpler problem, where the gauge
fields $A$ are taken to be abelian and one introduces certain new degrees
of freedom -- a spinor $\psi$, which satisfies the massless Dirac
equations $\Gamma_{\mu}D^{\mu}_{A}\psi=0$. The couplings between $A$ and
$\psi$ are described by, in addtion to the Dirac eqautions, a non-linear
relation $F^{+}(A)={\cal Q}(\psi,\bar{\psi})$, where $\cal Q$ is some
quadratic form in $\psi$, taking values in the anti self-dual part
$\Lambda^{2}_{+}$ of two-forms. This is the basic setup of
the Seiberg-Witten theory \cite{W94}. Now a natural question arises as
whether we can find an 8-dimensional analog of such a theory.
\vskip 1.2mm
\underbar{Manifolds with $Spin(7)$ Holonomy}. On 8-dimensional
manifold $X$ with $Spin(7)$ holonomy, there also exists a natural quadratic
form ${\cal Q}(\psi,\bar{\psi})$ valued in $\Lambda^{2}_{+}$. Indeed, given a
complex line bundle $\cal L$ and a spinor field $\psi\in S^{+}\otimes{\cal L}$,
one can construct a two-form $\zeta^{T}\Gamma_{\mu\nu}\psi=-\psi^{T}\Gamma_{
\mu\nu}\zeta$ and, according to \cite{AFSO}, it takes values in $\Lambda^{2}_{+}
\otimes{\cal L}$. One can also form the inner product $\bar{\psi}^{T}\zeta\in
\Lambda^{0}\otimes{\cal L}^{-1}$. It follows that the quadratic form ${\cal
Q}_{\mu\nu}(\psi,\bar{\psi})=(\bar{\psi}^{T}\zeta)(\zeta^{T}\Gamma_{\mu\nu}
\psi)$ belongs to $(\Lambda^{0}\otimes{\cal L}^{-1})\otimes(\Lambda^{2}_{+}
\otimes{\cal L})\cong\Lambda^{2}_{+}\otimes\mbb C$. Thus, by choosing a
unitary connection $A$ of $\cal L$, it is possible to write down an
8-dimensional analog of the Seiberg-Witten equations
\begin{equation}
\begin{array}{lll}
{\displaystyle F_{\mu\nu}^{+}(A)} & = &
{\displaystyle ia\cdot \Re
\left[(\bar{\psi}^{T}\zeta)(\zeta^{T}\Gamma_{\mu\nu}\psi)\right]
+ib\cdot\Im\left[(\bar{\psi}^{T}\zeta)(\zeta^{T}
\Gamma_{\mu\nu}\psi)\right],}\vspace{.3cm}\\
{\displaystyle \Gamma_{\mu}D_{A}^{\mu}\psi} & = & 0
\end{array}
\label{SW-1}
\end{equation}
where $a$, $b$ are real constants.

In order to see that both of the real part and the imaginary part of $\cal Q$
are not necessarily vanishing for generic $\psi\in S^{+}\otimes{\cal L}$,
one may work out $(\bar{\psi}^{T}\zeta)(\zeta^{T}\Gamma_{\mu\nu}\psi)$ in a
fully explicit form. Using the $\gamma$-matrices given in the Appendix we
find
$$
\zeta^{T}\Gamma_{12}\psi=-\zeta_{1}\psi_{2}+\zeta_{2}\psi_{1}
+\zeta_{3}\psi_{4}-\zeta_{4}\psi_{3}-\zeta_{5}\psi_{6}+
\zeta_{6}\psi_{5}+\zeta_{7}\psi_{8}-\zeta_{8}\psi_{7},
$$
$$\zeta^{T}\Gamma_{13}\psi=\zeta_{1}\psi_{4}-\zeta_{2}\psi_{3}
+\zeta_{3}\psi_{2}-\zeta_{4}\psi_{1}+\zeta_{5}\psi_{8}-
\zeta_{6}\psi_{7}+\zeta_{7}\psi_{6}-\zeta_{8}\psi_{5},\;\;\;
$$
$$\zeta^{T}\Gamma_{14}\psi=-\zeta_{1}\psi_{5}+\zeta_{2}\psi_{6}
-\zeta_{3}\psi_{7}+\zeta_{4}\psi_{8}+\zeta_{5}\psi_{1}-
\zeta_{6}\psi_{2}+\zeta_{7}\psi_{3}-\zeta_{8}\psi_{4},
$$
$$\zeta^{T}\Gamma_{15}\psi=\zeta_{1}\psi_{6}+\zeta_{2}\psi_{5}
+\zeta_{3}\psi_{8}+\zeta_{4}\psi_{7}-\zeta_{5}\psi_{2}-
\zeta_{6}\psi_{1}-\zeta_{7}\psi_{4}-\zeta_{8}\psi_{3},\;\;\;
$$
$$\zeta^{T}\Gamma_{16}\psi=-\zeta_{1}\psi_{3}-\zeta_{2}\psi_{4}
+\zeta_{3}\psi_{1}+\zeta_{4}\psi_{2}+\zeta_{5}\psi_{7}+
\zeta_{6}\psi_{8}-\zeta_{7}\psi_{5}-\zeta_{8}\psi_{6},
$$
$$\zeta^{T}\Gamma_{17}\psi=\zeta_{1}\psi_{7}+\zeta_{2}\psi_{8}
-\zeta_{3}\psi_{5}-\zeta_{4}\psi_{6}+\zeta_{5}\psi_{3}+
\zeta_{6}\psi_{4}-\zeta_{7}\psi_{1}-\zeta_{8}\psi_{2},\;\;\;
$$
$$\zeta^{T}\Gamma_{18}\psi=-\zeta_{1}\psi_{8}+\zeta_{2}\psi_{7}
+\zeta_{3}\psi_{6}-\zeta_{4}\psi_{5}+\zeta_{5}\psi_{4}-
\zeta_{6}\psi_{3}-\zeta_{7}\psi_{2}+\zeta_{8}\psi_{1},
$$
$$\zeta^{T}\Gamma_{23}\psi=\zeta_{1}\psi_{3}+\zeta_{2}\psi_{4}
-\zeta_{3}\psi_{1}-\zeta_{4}\psi_{2}+\zeta_{5}\psi_{7}+
\zeta_{6}\psi_{8}-\zeta_{7}\psi_{5}-\zeta_{8}\psi_{6},\;\;\;
$$
$$\zeta^{T}\Gamma_{24}\psi=-\zeta_{1}\psi_{6}-\zeta_{2}\psi_{5}
+\zeta_{3}\psi_{8}+\zeta_{4}\psi_{7}+\zeta_{5}\psi_{2}+
\zeta_{6}\psi_{1}-\zeta_{7}\psi_{4}-\zeta_{8}\psi_{3},
$$
$$\zeta^{T}\Gamma_{25}\psi=-\zeta_{1}\psi_{5}+\zeta_{2}\psi_{6}
+\zeta_{3}\psi_{7}-\zeta_{4}\psi_{8}+\zeta_{5}\psi_{1}-
\zeta_{6}\psi_{2}-\zeta_{7}\psi_{3}+\zeta_{8}\psi_{4},
$$
$$\zeta^{T}\Gamma_{26}\psi=\zeta_{1}\psi_{4}-\zeta_{2}\psi_{3}
+\zeta_{3}\psi_{2}-\zeta_{4}\psi_{1}-\zeta_{5}\psi_{8}+
\zeta_{6}\psi_{7}-\zeta_{7}\psi_{6}+\zeta_{8}\psi_{5},\;\;\;
$$
$$\zeta^{T}\Gamma_{27}\psi=-\zeta_{1}\psi_{8}+\zeta_{2}\psi_{7}
-\zeta_{3}\psi_{6}+\zeta_{4}\psi_{5}-\zeta_{5}\psi_{4}+
\zeta_{6}\psi_{3}-\zeta_{7}\psi_{2}+\zeta_{8}\psi_{1},
$$
$$\zeta^{T}\Gamma_{28}\psi=-\zeta_{1}\psi_{7}-\zeta_{2}\psi_{8}
+\zeta_{3}\psi_{5}-\zeta_{4}\psi_{6}+\zeta_{5}\psi_{3}+
\zeta_{6}\psi_{4}+\zeta_{7}\psi_{1}+\zeta_{8}\psi_{2},
$$
$$\zeta^{T}\Gamma_{34}\psi=\zeta_{1}\psi_{8}+\zeta_{2}\psi_{7}
+\zeta_{3}\psi_{6}+\zeta_{4}\psi_{5}-\zeta_{5}\psi_{4}-
\zeta_{6}\psi_{3}-\zeta_{7}\psi_{2}-\zeta_{8}\psi_{1},\;\;\;
$$
$$\zeta^{T}\Gamma_{35}\psi=\zeta_{1}\psi_{7}-\zeta_{2}\psi_{8}
+\zeta_{3}\psi_{5}-\zeta_{4}\psi_{6}-\zeta_{5}\psi_{3}+
\zeta_{6}\psi_{4}-\zeta_{7}\psi_{1}+\zeta_{8}\psi_{2},\;\;\;
$$
$$\zeta^{T}\Gamma_{36}\psi=\zeta_{1}\psi_{2}-\zeta_{2}\psi_{1}
-\zeta_{3}\psi_{4}+\zeta_{4}\psi_{3}-\zeta_{5}\psi_{6}+
\zeta_{6}\psi_{5}+\zeta_{7}\psi_{8}-\zeta_{8}\psi_{7},\;\;\;
$$
$$\zeta^{T}\Gamma_{37}\psi=-\zeta_{1}\psi_{6}+\zeta_{2}\psi_{5}
+\zeta_{3}\psi_{8}-\zeta_{4}\psi_{7}-\zeta_{5}\psi_{2}+
\zeta_{6}\psi_{1}+\zeta_{7}\psi_{4}-\zeta_{8}\psi_{3},
$$
$$\zeta^{T}\Gamma_{38}\psi=-\zeta_{1}\psi_{5}-\zeta_{2}\psi_{6}
+\zeta_{3}\psi_{7}+\zeta_{4}\psi_{8}+\zeta_{5}\psi_{1}+
\zeta_{6}\psi_{2}-\zeta_{7}\psi_{3}-\zeta_{8}\psi_{4},
$$
$$\zeta^{T}\Gamma_{45}\psi=\zeta_{1}\psi_{2}-\zeta_{2}\psi_{1}
+\zeta_{3}\psi_{4}-\zeta_{4}\psi_{3}+\zeta_{5}\psi_{6}-
\zeta_{6}\psi_{5}+\zeta_{7}\psi_{8}-\zeta_{8}\psi_{7},\;\;\;
$$
$$\zeta^{T}\Gamma_{46}\psi=-\zeta_{1}\psi_{7}+\zeta_{2}\psi_{8}
+\zeta_{3}\psi_{5}-\zeta_{4}\psi_{6}-\zeta_{5}\psi_{3}+
\zeta_{6}\psi_{4}+\zeta_{7}\psi_{1}-\zeta_{8}\psi_{2},
$$
$$\zeta^{T}\Gamma_{47}\psi=-\zeta_{1}\psi_{3}+\zeta_{2}\psi_{4}
+\zeta_{3}\psi_{1}-\zeta_{4}\psi_{2}+\zeta_{5}\psi_{7}-
\zeta_{6}\psi_{8}-\zeta_{7}\psi_{5}+\zeta_{8}\psi_{6},
$$
$$\zeta^{T}\Gamma_{48}\psi=-\zeta_{1}\psi_{4}-\zeta_{2}\psi_{3}
+\zeta_{3}\psi_{2}+\zeta_{4}\psi_{1}-\zeta_{5}\psi_{8}-
\zeta_{6}\psi_{7}+\zeta_{7}\psi_{6}+\zeta_{8}\psi_{5},
$$
$$\zeta^{T}\Gamma_{56}\psi=\zeta_{1}\psi_{8}+\zeta_{2}\psi_{7}
-\zeta_{3}\psi_{6}-\zeta_{4}\psi_{5}+\zeta_{5}\psi_{4}+
\zeta_{6}\psi_{3}-\zeta_{7}\psi_{2}-\zeta_{8}\psi_{1},\;\;\;
$$
$$\zeta^{T}\Gamma_{57}\psi=\zeta_{1}\psi_{4}+\zeta_{2}\psi_{3}
-\zeta_{3}\psi_{2}-\zeta_{4}\psi_{1}-\zeta_{5}\psi_{8}-
\zeta_{6}\psi_{7}+\zeta_{7}\psi_{6}+\zeta_{8}\psi_{5},\;\;\;
$$
$$\zeta^{T}\Gamma_{58}\psi=-\zeta_{1}\psi_{3}+\zeta_{2}\psi_{4}
+\zeta_{3}\psi_{1}-\zeta_{4}\psi_{2}-\zeta_{5}\psi_{7}+
\zeta_{6}\psi_{8}+\zeta_{7}\psi_{5}-\zeta_{8}\psi_{6},
$$
$$\zeta^{T}\Gamma_{67}\psi=\zeta_{1}\psi_{5}+\zeta_{2}\psi_{6}
+\zeta_{3}\psi_{7}+\zeta_{4}\psi_{8}-\zeta_{5}\psi_{1}-
\zeta_{6}\psi_{2}-\zeta_{7}\psi_{3}-\zeta_{8}\psi_{4},\;\;\;
$$
$$\zeta^{T}\Gamma_{68}\psi=-\zeta_{1}\psi_{6}+\zeta_{2}\psi_{5}
-\zeta_{3}\psi_{8}+\zeta_{4}\psi_{7}-\zeta_{5}\psi_{2}+
\zeta_{6}\psi_{1}-\zeta_{7}\psi_{4}+\zeta_{8}\psi_{3},
$$
$$\zeta^{T}\Gamma_{78}\psi=-\zeta_{1}\psi_{2}+\zeta_{2}\psi_{1}
-\zeta_{3}\psi_{4}+\zeta_{4}\psi_{3}+\zeta_{5}\psi_{6}-
\zeta_{6}\psi_{5}+\zeta_{7}\psi_{8}-\zeta_{8}\psi_{7}.
$$
So we get, for example,
\begin{equation}
(\bar{\psi}^{T}\zeta)(\zeta^{T}\Gamma_{12}\psi)=\left(\sum_{A=1}^{8}
\bar{\psi}_{A}\zeta_{A}\right)\cdot\sum_{a=1}^{4}(-1)^{a}\left(
\zeta_{2a-1}\psi_{2a}-\zeta_{2a}\psi_{2a-1}\right).
\label{explicit}
\end{equation}
If we write $\psi=\chi+i\eta$, then the real and imaginary parts of
(\ref{explicit}) read
\begin{equation}
\begin{array}{lll}
\displaystyle{
\Re\left[(\bar{\psi}^{T}\zeta)(\zeta^{T}\Gamma_{12}\psi)\right]} & = &
\displaystyle{
\left(\sum_{A=1}^{8}\chi_{A}\zeta_{A}\right)\cdot\sum_{a=1}^{4}(-1)^{a}
\left(\zeta_{2a-1}\chi_{2a}-\zeta_{2a}\chi_{2a-1}\right)}\vspace{.3cm}\\
& + & \displaystyle{
\left(\sum_{A=1}^{8}\eta_{A}\zeta_{A}\right)\cdot\sum_{a=1}^{4}(-1)^{a}
\left(\zeta_{2a-1}\eta_{2a}-\zeta_{2a}\eta_{2a-1}\right)},
\end{array}
\label{explicitRe}
\end{equation}
\begin{equation}
\begin{array}{lll}
\displaystyle{
\Im\left[(\bar{\psi}^{T}\zeta)(\zeta^{T}\Gamma_{12}\psi)\right]} & = &
\displaystyle{
\left(\sum_{A=1}^{8}\chi_{A}\zeta_{A}\right)\cdot\sum_{a=1}^{4}(-1)^{a}
\left(\zeta_{2a-1}\eta_{2a}-\zeta_{2a}\eta_{2a-1}\right)}\vspace{.3cm}\\
& - & \displaystyle{
\left(\sum_{A=1}^{8}\eta_{A}\zeta_{A}\right)\cdot\sum_{a=1}^{4}(-1)^{a}
\left(\zeta_{2a-1}\chi_{2a}-\zeta_{2a}\chi_{2a-1}\right)}.
\end{array}
\label{explicitIm}
\end{equation}
Other $(\mu,\nu)$-components can be written down similarly.

The above explicit result shows that for generic spinors $\psi$, both the real
part and the imaginary part of ${\cal Q}_{\mu\nu}(\psi,\bar{\psi})$ are indeed
not zero. This is different from the 4-dimensional Seiberg-Witten theory,
where the quadratic form ${\cal Q}_{\mu\nu}(\psi,\bar{\psi})=\bar{\psi}^{T}
\Gamma_{\mu\nu}\psi$ is essentially purely imaginary, as we can choose the
$Spin(4)$ Lie algebra generators $\Gamma_{\mu\nu}$ to be anti-hermitian. The
difference stems from the fact that in 4-dimensional theory the quadratic
takes the ``diagonal form'' $\bar{\psi}^{T}\Gamma_{\mu\nu}\psi\in\Lambda_{
+}^{2}$ while in eight dimensions, such a diagonal form does not belong to
$\Lambda_{+}^{2}$ (though it is still purely imaginary). In order to define a
reasonable ${\cal Q}\in\Lambda_{+}^{2}$ in 8 dimensions, we have to
decompose the spinor $\psi\in S^{+}\cong {\bf 1}\oplus{\bf 7}$ into two
parts $\psi=\psi_{\bf 1}+\psi_{\bf 7}$, one of which, $\psi_{\bf 1}\equiv
(\zeta^{T}\psi)\zeta$, is in $\bf 1$, {\it i.e.} the trivial module of
$Spin(7)$, and the other of which, $\psi_{\bf 7}$, belongs to $\bf 7$,
namely the seven-dimensional irreducible module of $Spin(7)$. Since this
decomposition
is orthogonal and since $\Gamma_{\mu\nu}\zeta\in{\bf 7}$, we have 
$\bar{\psi}^{T}\zeta=\bar{\psi}_{\bf 1}^{T}\zeta$ and
$\zeta^{T}\Gamma_{\mu\nu}\psi=\zeta^{T}\Gamma_{\mu\nu}\psi_{\bf 7}$. Thus,
the quadratic form ${\cal Q}_{\mu\nu}=(\bar{\psi}^{T}\zeta)(\zeta^{T}\Gamma_{
\mu\nu}\psi)=(\bar{\psi}_{\bf 1}^{T}\zeta)(\zeta^{T}\Gamma_{\mu\nu}\psi_{\bf
7})$ we have just constructed is really an ``off-diagonal'' product between
the independent degrees of freedom $\psi_{\bf 1}$ and $\psi_{\bf 7}$. Such a
product cannot be automatically real or purely imaginary. This explains why
in the first equation of (\ref{SW-1}), we have splitted the quadratic form
into its real and imaginary parts, and introduced two real coefficients $a$
and $b$.

There is a more compact way to write down the real and imaginary parts of
$\cal Q$:
\begin{equation}
\begin{array}{lll}
\Re\left[(\bar{\psi}^{T}\zeta)(\zeta^{T}\Gamma_{\mu\nu}\psi)\right]
&=& \displaystyle{\frac{1}{2\cdot 4!}{\Omega^{\lambda\rho\sigma}}_{[\mu}
(\psi^{\dag}\Gamma_{\nu]\lambda\rho\sigma}\psi)},\vspace{.3cm}\\
i\Im\left[(\bar{\psi}^{T}\zeta)(\zeta^{T}\Gamma_{\mu\nu}\psi)\right]
& = & \displaystyle{\frac{1}{8}(\psi^{\dag}\Gamma_{\mu\nu}\psi)-\frac{1}{16}
\Omega_{\mu\nu\alpha
\beta}(\psi^{\dag}\Gamma^{\alpha\beta}\psi)}\vspace{.3cm}\\
& \equiv & \displaystyle{\frac{1}{2}{(P^{+})_{\mu\nu}}^{\alpha\beta}(\psi^{\dag}
\Gamma_{\alpha\beta}\psi)}
\end{array}
\label{ReIm}
\end{equation}
where $P^{+}:\Lambda^{2}\rightarrow\Lambda^{2}_{+}$ is the orthorgonal
projection \cite{AFSO} of two-forms onto $\Lambda^{2}_{+}$. Note that the
imaginary part of $\cal Q$ resembles the term $\psi^{\dag}\Gamma_{\mu\nu}\psi$
in the Seiberg-Witten theory, but in eight dimensions there are additional
ingredients in the construction of a general quadratic form valued in
$\Lambda^{2}_{+}$: we have terms involving ${\Omega^{\lambda\rho\sigma}}_{[\mu}
\Gamma_{\nu]\lambda\rho\sigma}$. Such terms are forbidden in 4
dimensions since there ${\Omega^{\lambda\rho\sigma}}_{\mu}\propto
{\epsilon^{\lambda\rho\sigma}}_{\mu}$, $\Gamma_{\nu\lambda\rho\sigma}
\propto\epsilon_{\nu\lambda\rho\sigma}$ and ${\epsilon^{\lambda\rho
\sigma}}_{[\mu}\epsilon_{\nu]\lambda\rho\sigma}=0$.

Let us discuss another difference between the 8-dimensional and 4-dimensional
theories. Writing down the equations in such theories requires to fix
certain geometrical data on the underlying manifold. For example, in order to construct
the anti self-dual part $F^{+}$ of the curvature tensor in the 4-dimensional
theory, one has to pick up a Hodge star operator, whose definition depends on the
conforml structure of the manifold. Thus, the geometrical data -- a conformal
structure of the 4-manifold -- enters natually in the first Seiberg-Witten
equation $F^{+}_{\mu\nu}\sim{\cal Q}_{\mu\nu}$. Such geometrical data also
enters in the the second Seiberg-Witten equation, {\it i.e.} the massless
Dirac equation in 4 dimensions, as that equation is conformally invariant and
it also depends on the choice of a conformal structure. In the 8-dimensional
theory, the construction of the first equation involves another data,
$\Omega$, which is the $Spin(7)$-invariant 4-form calibrating the
underlying geometry. This can be expected, since as long as the
self-duality structures are concerned $\Omega$ will play a role similar to the
Hodge star operator in 4 dimensions. What makes the 8-dimensional theory
different from that in 4 dimensions is that the geometrical data $\Omega$
does not enter in the Dirac equation. Thus, it should not be very suprising
when we find that the functional formalism of (\ref{SW-1}) in general does
not allow the delicate cancellations as in the 4-dimensional theory. In
particular, we do not know at present how to handle the uncancelled terms
involving $F^{-}$, arsing from the functional $||\Gamma_{\mu}D_{A}^{\mu}
\psi||^{2}$ of the Dirac equation. One possible resolusion is to modify the
second equation in (\ref{SW-1}) so that it depends on the form $\Omega$
(through the $Spin(7)$-invariant spinor $\zeta$).
\vskip 1.2mm
\underbar{Manifolds with $SU(4)$ Holonomy}.
Now we try to formulate an eight dimensional analog of the Seiberg-Witten
equations on manifolds with the $SU(4)$ holonomy group. The starting point
will be similar to that in the $Spin(7)$ case: One wishes to replace the
nonabelian instanton equation $F^{(0,2)}_{+\bar{i}\bar{j}}=0$ by an
abelian, monopole-like equation $F^{(0,2)}_{+\bar{i}\bar{j}}={\cal Q}_{\bar{i}\bar{j}}
(\psi,\bar{\psi})$, where $\psi\in S^{+}_{c}\otimes{\cal L}$ is a spinor
field twisted by some complex line bundle $\cal L$, and ${\cal Q}_{\bar{i}
\bar{j}}(\psi,\bar{\psi})$ denotes a certain quadratic form valued in
$\Lambda^{0,2}_{+}$. Our first task is thus to find out such a quadratic form.

The condition for a (0,2)-form $\beta_{\bar{i}\bar{j}}$ to be valued in
$\Lambda^{0,2}_{+}$ is that it obeys the eigenvalue equation $*_{\Omega}\beta
=-4\beta$. So according to (\ref{Ostar}), $\beta\in\Lambda^{0,2}_{+}$ is
characterized by the equations
\begin{equation}
\bar{\Omega}_{\bar{i}\bar{j}\bar{k}\bar{l}}\bar{\beta}^{\bar{k}\bar{l}}
=-8\beta_{\bar{i}\bar{j}}\quad\Longleftrightarrow\quad
\Omega_{ijkl}\beta^{kl}=-8\bar{\beta}_{ij}.
\label{F-=0}
\end{equation}
Our key observation here is that the spinor $\gamma^{ij}\bar{\zeta}$
satisfies an equation with the same structure as the second one in
(\ref{F-=0}). To see this, multiplying (\ref{fierz3-2}) by $\zeta^{T}
\gamma_{ijkl}$ from the left, we find
$$
\Omega_{ijkl}\bar{\zeta}=\frac{1}{8}\zeta^{T}(\gamma_{ijkl}+g^{m\bar{n}}
\gamma_{ijkl}\gamma_{m\bar{n}})+\frac{1}{32}\zeta^{T}g^{m\bar{s}}g^{n\bar{r}}
\gamma_{ijkl}\gamma_{mn\bar{r}\bar{s}}.
$$
Notice that $\gamma_{i_{1}\cdots i_{p}}=\gamma_{i_{1}}\cdots\gamma_{i_{p}}$
and $\gamma_{i_{1}\cdots i_{p}\bar{j}_{1}\cdots\bar{j}_{q}}=0$ for $p>4$. So
one can use Eq.(\ref{gammaID-C}) repeatedly to compute $g^{m\bar{n}}
\gamma_{ijkl}\gamma_{m\bar{n}}$ as well as $g^{m\bar{s}}g^{n\bar{r}}
\gamma_{ijkl}\gamma_{mn\bar{r}\bar{s}}$, and the result simply reads
$$
g^{m\bar{n}}\gamma_{ijkl}\gamma_{m\bar{n}}=4\gamma_{ijkl},\quad
g^{m\bar{s}}g^{n\bar{r}}\gamma_{ijkl}\gamma_{mn\bar{r}\bar{s}}=
12\gamma_{ijkl}.
$$
Consequently, we have \cite{MMMS}
\begin{equation}
\Omega_{ijkl}\bar{\zeta}
=\zeta^{T}\gamma_{ijkl}=\gamma_{ijkl}\zeta
\label{tmp}
\end{equation}
(where the last identity comes from the symmetric property of the matrix
$\gamma_{ijkl}$). Now with the help of (\ref{tmp}) and (\ref{gammaID-C}), we
can do some further computations:
$$\Omega_{ijkl}\gamma^{kl}\bar{\zeta}=\gamma^{kl}(\Omega_{ijkl}\bar{\zeta})
=\gamma^{kl}\gamma_{ijkl}\zeta=g^{k\bar{m}}g^{l\bar{n}}\gamma_{\bar{m}\bar{n}}
\gamma_{ijkl}\zeta
$$
$$
=g^{k\bar{m}}g^{l\bar{n}}\gamma_{\bar{m}}\gamma_{ijkl\bar{n}}\zeta
+g^{k\bar{m}}\gamma_{\bar{m}}\gamma_{ijk}\zeta\quad\;\quad\quad\quad\quad\quad\quad
$$
$$
=g^{k\bar{m}}g^{l\bar{n}}\gamma_{ijkl
\bar{m}\bar{n}}\zeta-2g^{k\bar{m}}\gamma_{ijk\bar{m}}\zeta-2\gamma_{ij}\zeta
=-8\gamma_{ij}\zeta.
$$
So finally we arrive at 
\begin{equation}
\Omega_{ijkl}\gamma^{kl}\bar{\zeta}=-8\gamma_{ij}\zeta,
\label{zetaID}
\end{equation}
which shows that $\gamma^{ij}\bar{\zeta}$ has the same tensor properties as
an anti self-dual two-form $\beta^{ij}\in\Lambda^{0,2}_{+}$.

Thus, given any spinor $\psi\in S^{+}_{c}\otimes{\cal L}$, one can use the
isomorphism (\ref{isomComplex1}) to construct a form $\beta^{ij}=-\psi^{T}
\gamma^{ij}\bar{\zeta}=\zeta^{\dag}\gamma^{ij}\psi\in (S^{+}_{c}\otimes{\cal
L})\otimes\mbb C\cong \Lambda^{0,{\rm even}}\otimes {\cal L}$. Naively, the
identity (\ref{zetaID}) indicates that such a form should obey the anti
self-duality equation (\ref{F-=0}), and thus it would belong to the
subspace $\Lambda^{0,2}_{+}\otimes{\cal L}$:
\begin{equation}
\beta^{ij}\equiv\zeta^{\dag}\gamma^{ij}\psi\in\Lambda^{0,2}_{+}\otimes
{\cal L}.
\label{beta}
\end{equation}
To construct a quadratic form ${\cal Q}_{\bar{i}\bar{j}}(\psi,\bar{\psi})\in
\Lambda^{0,2}_{+}$, one still needs another form $\alpha$, which should be
anti-linear in $\psi$ and valued in $\Lambda^{0,0}\otimes{\cal L}^{-1}$, so
that the factor $\cal L$ could be cancelled when forming the product $\alpha
\beta_{\bar{i}\bar{j}}$. The simplest choice of such a form would be
\begin{equation}
\alpha\equiv\zeta^{T}\bar{\psi}=\psi^{\dag}\zeta\in\Lambda^{0,0}
\otimes{\cal L}^{-1}.
\label{alpha}
\end{equation}
So at first sight we expect that the quadratic form we are seeking should
look like\footnote{(\ref{quadratic}) is
very similar to the quadratic form ${\cal Q}_{\mu\nu}$
constructed in the real case. One may decompose $\psi$ into
$\psi=\psi_{\parallel}+\psi_{\perp}$, where $\psi_{\parallel}$ is valued in
the $SU(4)$ invariant subspace of $S^{+}_{c}$ spanned by $\zeta$ and
$\bar{\zeta}$, and $\psi_{\perp}$ lives in the subspace orthogonal to it.
Using the facts that $\zeta^{T}\zeta=0$, $\zeta^{\dag}\gamma_{\bar{i}\bar{j}}
\zeta=\zeta^{\dag}\gamma_{\bar{i}\bar{j}}\bar{\zeta}=0$, we find that
(\ref{quadratic}) can be represented as an ``off-diagonal'' product of the
two independent degrees of freedom $\psi_{\parallel}$ and $\psi_{\perp}$,
namely ${\cal Q}_{\bar{i}\bar{j}}=(\psi_{\parallel}^{\dag}\zeta)(\zeta^{\dag}
\gamma_{\bar{i}\bar{j}}\psi_{\perp})$. This resembles the $Spin(7)$ case,
where the quadratic form is also an off-diagonal product of two independent
degrees of freedom.}:
\begin{equation}
{\cal Q}_{\bar{i}\bar{j}}(\psi,\bar{\psi})\sim\alpha\beta_{\bar{i}\bar{j}}
=(\psi^{\dag}\zeta)(\zeta^{\dag}\gamma_{\bar{i}\bar{j}}\psi).
\label{quadratic}
\end{equation}

However, there is a subtlety in the above construction, which appears only in
the complex case. In our definition of self-duality, the star operator $*_{
\Omega}$ given in (\ref{Ostar}) is conjugate-linear rather than linear. Thus,
even if $\beta_{\bar{i}\bar{j}}$ is anti self-dual, namely it obeys the
condition (\ref{F-=0}), the quantity $\alpha\beta_{\bar{i}\bar{j}}$ needs not
to be such a form for {\it complex} $\alpha\in\Lambda^{0,0}\otimes{\cal L}^{
-1}$. We cannot simply take $\alpha$ to be real as $\cal L$ should be a
nontrivial complex line bundle. Moreover, since $\psi$ is also a complex
spinor, the (0,2)-form $\beta^{ij}=-\psi^{T}\gamma^{ij}\bar{\zeta}$ defined
by (\ref{beta}) is not really valued in $\Lambda^{0,2}_{+}$, even though
$\gamma^{ij}\bar{\zeta}$ behaves as an anti self-dual tensor. To solve this
problem, let us introduce a pair $\alpha$, $\alpha'$ of (0,0)-forms as well
as a pair $\beta_{\bar{i}\bar{j}}$, $\beta'_{\bar{i}\bar{j}}$ of (0,2)-forms,
specified as
\begin{equation}
\begin{array}{lll}
\alpha & = & \psi^{\dag}\zeta\in\Lambda^{0,0}\otimes{\cal L}^{-1},\quad
\alpha'=\bar{\alpha}=\zeta^{\dag}\psi\in\Lambda^{0,0}\otimes{\cal L}\vspace{.3cm}\\
\beta^{ij} &=& \zeta^{\dag}\gamma^{ij}\psi\in\Lambda^{0,2}\otimes{\cal L},
\quad\beta'^{ij} = \zeta^{\dag}\gamma^{ij}\bar{\psi}\in\Lambda^{0,2}\otimes
{\cal L}^{-1},
\end{array}
\label{alphabeta}
\end{equation}
and construct the product
\begin{equation}
{\cal Q}_{\bar{i}\bar{j}}(\psi,\bar{\psi})\equiv c\alpha\beta_{\bar{i}\bar{j}}
+\bar{c}\alpha'\beta'_{\bar{i}\bar{j}}=c(\psi^{\dag}\zeta)(\zeta^{\dag}\gamma_{
\bar{i}\bar{j}}\psi)+\bar{c}(\psi^{T}\bar{\zeta})(\zeta^{\dag}\gamma_{\bar{i}\bar{j}}
\bar{\psi})
\label{quadratic'}
\end{equation}
with $c\in\mbb C$ being an arbitrarily fixed complex number (similar to the
real numbers $a$, $b$ in the $Spin(7)$ case). One then uses (\ref{zetaID}) to
derive
$$
\Omega_{ijkl}{\cal Q}^{kl}=-c(\psi^{\dag}\zeta)(\psi^{T}\Omega_{ijkl}
\gamma^{kl}\bar{\zeta})-\bar{c}(\psi^{T}\bar{\zeta})(\psi^{\dag}\Omega_{ijkl}
\gamma^{kl}\bar{\zeta})
$$
$$
=8c(\psi^{\dag}\zeta)(\psi^{T}\gamma_{ij}\zeta)+8\bar{c}(\psi^{T}\bar{\zeta})
(\psi^{\dag}\gamma_{ij}\zeta)=-8\bar{\cal Q}_{ij},
$$
which indicates that ${\cal Q}_{\bar{i}\bar{j}}$ is now in
$\Lambda^{0,2}_{+}$.

Having constructed a quadratic form ${\cal Q}_{\bar{i}\bar{j}}$ with the
right properties, we can immediately write down the first equation analogous
to Seiberg and Witten's:
\begin{equation}
F^{(0,2)}_{+\bar{i}\bar{j}}=c\alpha\beta_{\bar{i}\bar{j}}+\bar{c}\alpha'
\beta'_{\bar{i}\bar{j}}=c(\psi^{\dag}\zeta)(\zeta^{\dag}\gamma_{\bar{i}
\bar{j}}\psi)+\bar{c}(\psi^{\dag}\gamma_{\bar{i}\bar{j}}\bar{\zeta})
(\zeta^{\dag}\psi).
\label{SW-1-C}
\end{equation}
One may also write down a similar equation for $F^{(2,0)}_{+ij}$ by taking
the complex conjugate of (\ref{SW-1-C}). It should be pointed out, however,
that at this stage we have not yet established another kind of equation
(something like $F^{(1,1)}_{\omega}\sim -\frac{1}{2}\omega(||\alpha||^{2}-
||\beta||^{2})$ as in the 4-dimensional theory), which governs the
(1,1)-component of $F$ in the direction along the K\"{a}ller form $\omega_{i
\bar{j}}=ig_{i\bar{j}}$. To obtain such an equation, one
should use a new star operator $*_{\Theta}$ with $\Theta\sim \Re(\Omega)
+\omega^{2}$ to define self-duality \cite{tian}.

Next we consider the Dirac equation $D_{A}\psi=0$. With the isomorphism $(
S^{+}_{c},S^{-}_{c})\otimes\mbb C\cong (\Lambda^{0,{\rm even}},\Lambda^{0,{
\rm odd}})$, the twisted Dirac operator $D_{A}:S^{+}_{c}\otimes{\cal L}
\rightarrow S^{-}_{c}\otimes{\cal L}$ becomes $D_{A}=\bar{\partial}_{A}+\bar{
\partial}^{*}_{A}:\Lambda^{0,{\rm even}}\otimes{\cal L}\rightarrow\Lambda^{0,
{\rm odd}}\otimes{\cal L}$. Let us restrict this operator to $(\Lambda^{0,0}
\oplus\Lambda^{0,2})\otimes{\cal L}$. Under this restriction, the spinor
$\psi\in\Lambda^{0,{\rm even}}\otimes{\cal L}$ has the components $\alpha'=
\bar{\alpha}\in\Lambda^{0,0}\otimes{\cal L}$ and $\beta\in\Lambda^{0,2}
\otimes{\cal L}$, and the Dirac equation is reduced simply to
\begin{equation}
\bar{\partial}_{A}\bar{\alpha}+\bar{\partial}_{A}^{*}\beta=0.
\label{SW-2-C}
\end{equation}
This constitutes the second equation in our theory.

Although Eq.(\ref{SW-1-C})-(\ref{SW-2-C}) resemble the four-dimensional
Seiberg-Witten equations on K\"{a}hler manifolds \cite{W94}, it should be
pointed out that here the Dirac equation (\ref{SW-2-C}) in general does not
allow a simple decomposition into $\bar{\partial}_{A}\bar{\alpha}=\bar{
\partial}_{A}^{*}\beta=0$. This makes a computation of the relevant
invariants quite difficult. This difficulty is related to a problem
appeared in the $Spin(7)$ case, where we mentioned that there is an
uncancelled term involving $F^{-}$ in the functional formalism.
\section{$S$-duality in Abelian Gauge Theory}
In this section we turn to the abelian gauge theory in eight dimensions. For
simplicity, we will consider only the case when $X$ has the holonomy group
$Spin(7)$. Classically we have a $U(1)$ gauge field $A_{\mu}$ and its
field strength $F_{\mu\nu}=\partial_{\mu}A_{\nu}-\partial_{\nu}A_{\mu}$,
together with the following action functional
\begin{equation}
\begin{array}{lll}
S[A] & = & \displaystyle\frac{1}{2g^{2}}\int_{X}F_{\mu\nu}F^{\mu\nu}
+\frac{i\theta}{8\pi^{2}}\int_{X}\Omega\wedge F\wedge F\vspace{.3cm}\\
& = & \displaystyle\frac{1}{2g^{2}}\int_{X}F_{\mu\nu}F^{\mu\nu}
+\frac{i\theta}{32\pi^{2}}\int_{X}\Omega_{\mu\nu\alpha\beta}
F^{\mu\nu}F^{\alpha\beta}.
\end{array}
\label{abelianAction}
\end{equation}
Given such data, the partition function $Z(g,\theta)$ can be formally defined
as the Euclidean path-integral
\begin{equation}
Z(g,\theta)=\int [dA]\,e^{-S[A]}.
\label{partitionFunc}
\end{equation}

Let us analyze the partition function in some detail. Usually, it is convenient
to change the integration variables $A\rightarrow F$. The routine is quite standard: Just as in four
dimensions, $F$ is not an independent variable, and it must be subject to the
Bianchi identity $dF=0$. If we write $dF=\frac{1}{2}\partial_{\lambda}F_{\mu
\nu}dx^{\lambda}\wedge dx^{\mu}\wedge dx^{\nu}$, then one easily deduces from
the closeness and self-duality of $\Omega$ that
$$
dF\wedge\Omega\wedge dx^{\nu}=\frac{1}{2}\partial_{\mu}[(*_{\Omega}F)^{\mu\nu}]\,
d({\rm vol}).
$$
This implies that the Bianchi identity $dF=0$ can be replaced by a constraint
\begin{equation}
\partial_{\nu}[(*_{\Omega}F)^{\mu\nu}]=0
\label{constaint}
\end{equation}
on the field strength. Consequently, the partition function
(\ref{partitionFunc}) has a path-integral representation over $F$, with the
delta function $\delta(\partial_{\nu}[(*_{\Omega}F)^{\mu\nu}])$ inserted.
Such a delta function can be written as another path-integral over some
auxiliary field $A^{D}_{\mu}$. Thus, one may write
\begin{equation}
\begin{array}{lll}
Z(g,\theta)&=&\displaystyle\int [dF][dA^{D}]\,e^{-S+i\int_{X}
A_{\mu}^{D}\partial_{\nu}(*_{\Omega}F)^{\mu\nu}}\vspace{.3cm}\\
&=&\displaystyle\int [dF][dA^{D}]\,e^{-S+
\frac{i}{2}\int_{X} F_{\mu\nu}^{D}(*_{\Omega}F)^{\mu\nu}},
\end{array}
\label{tmp1}
\end{equation}
where $F^{D}_{\mu\nu}=\partial_{\mu}A^{D}_{\nu}-\partial_{\nu}A^{D}_{\mu}$ is
the field strength of $A^{D}_{\mu}$. If we first integrate out the auxiliary
field $A^{D}$, then the resulting expressonis is nothing but $\int [dF]\delta
(\partial_{\nu}[(*_{\Omega}F)^{\mu\nu}])e^{-S}$, which is equivalent to the
original partition function (\ref{partitionFunc}).

Alternatively, one can integrate out $F$ in (\ref{tmp1}) first, leaving an
effective action for the auxiliary field, which governs the dynamics of the
collective variable $A^{D}_{\mu}$. To achieve this, let us decompose $F=F^{+}
+F^{-}$ into two independent components $F^{+}$, $F^{-}$ and write $[dF]=
[dF^{+}][dF^{-}]$. In terms of these components, we have
\begin{equation}
\begin{array}{lll}
S&=&\displaystyle\left(\frac{1}{2g^{2}}-i\frac{3\theta}{16\pi^{2}}\right)
\int_{X} F^{+}_{\mu\nu}F^{+\mu\nu}
+\left(\frac{1}{2g^{2}}+i\frac{\theta}{16\pi^{2}}\right)\int_{X}
F^{-}_{\mu\nu}F^{-\mu\nu}\vspace{.3cm}\\
&\equiv & \displaystyle\frac{3}{4e_{+}^{2}}\int_{X} F^{+}_{\mu\nu}F^{+\mu\nu}
+\frac{1}{4e_{-}^{2}}\int_{X} F^{-}_{\mu\nu}F^{-\mu\nu},
\end{array}
\label{tmp2}
\end{equation}
\begin{equation}
\frac{i}{2}\int_{X} F_{\mu\nu}^{D}(*_{\Omega}F)^{\mu\nu}=
-\frac{3i}{2}\int_{X} F_{\mu\nu}^{D+}F^{+\mu\nu}
+\frac{i}{2}\int_{X} F_{\mu\nu}^{D-}F^{-\mu\nu}.
\label{tmp3}
\end{equation}
So substituting (\ref{tmp2})-(\ref{tmp3}) into (\ref{tmp1}) yields a product
of two gaussian integrals over $F^{\pm}$. An explicit evaluation of these
integrals gives
\begin{equation}
\begin{array}{lll}
Z(g,\theta)&=&\displaystyle\int [dA^{D}]\,e^{-\widetilde{S}[A^{D}]},
\vspace{.3cm}\\
\widetilde{S}[A^{D}]&=&\displaystyle\frac{3e_{+}^{2}}{4}\int_{X} F^{D+}_{\mu\nu}F^{D+\mu\nu}
+\frac{e_{-}^{2}}{4}\int_{X} F^{D-}_{\mu\nu}F^{D-\mu\nu}.
\end{array}
\label{dualPartition}
\end{equation}
We thus obtain a dual description of the original theory using the collective
field $A^{D}$, in which the coupling constants get transformed:
\begin{equation}
e_{\pm}^{2}\rightarrow \widetilde{e}_{\pm}^{2}=\frac{1}{e_{\pm}^{2}}.
\label{inverse}
\end{equation}
Eq.(\ref{inverse}) characterizes a generalized $S$-duality in eight
dimensions.

The above discussion is somewhat rough and we ignored several subtleties
arising from regularization. In four dimensions, a more careful study
\cite{W95} shows that the partition function transforms as a modular form,
and this provides a precise test of the $S$-duality. When entering in
eight dimensions, however, one sees from (\ref{tmp2}) that the action does
not takes the form $S\propto i\int (\tau (F^{+})^{2}-\bar{\tau}(F^{-})^{2})$,
so the partition function $Z(g,\theta)$ will not be parametrized neatly by a
single complex coupling $\tau$ along with its conjugate $\bar{\tau}$; more
naturally, $Z(g,\theta)$ should be parametrized by $(e^{+},e^{-})$, and
$e^{\pm}$ are not complex conjugate to each other. It seems rather difficult
to write down a simple modular form expression for the partition function of
the eight-dimensional theory. Without such a modular form our understanding
of the generalized $S$-duality is quite incomplete.

\acknowledgments
We would like to thank I. Singer for some helpful conversations.
\appendix
\section{$\gamma$-Matrices and Clifford Calculus}
In the text we used $Cl(8)$ to denote the 8-dimensional Clifford algebra. This
algebra has a real, irreducible representation $\rho: Cl(8)\rightarrow
End(S)$. According to the standard argument, $\rho(Cl(8))$ constitutes the
algebra $\mbb R(16)$ of $16\times 16$ real matrices, acting on the
16-dimensional vector space $S\cong\mbb R^{16}$. The following isomorphism
between $Cl(8)$ and the wedge algebra $\wedge^{*}\mbb R^{8}$ is quite
evident:
\begin{equation}
Cl(8)\cong{\wedge}^{*}\,\mbb R^{8}.
\end{equation}
In particular, if we introduce a set of orthogonal generators of $Cl(8)$,
$e_{\mu}\in\wedge^{1}\mbb R^{8}$ ($1\leq\mu\leq 8$), with the rule of
Clifford multiplications
\begin{equation}
e_{\mu}\cdot e_{\nu}+e_{\nu}\cdot e_{\mu}=-2\,\langle e_{\mu},e_{\nu}
\rangle\equiv -2g_{\mu\nu},
\label{cliff}
\end{equation}
then the $p$-``form'' $e_{\mu_{1}}\wedge e_{\mu_{2}}\wedge
\cdots\wedge e_{\mu_{p}}\in\wedge^{p}\mbb R^{8}$ canonically has the
representation
\begin{equation}
e_{\mu_{1}}\wedge e_{\mu_{2}}\wedge\cdots\wedge e_{\mu_{p}}
\longleftrightarrow
\Gamma_{\mu_{1}\mu_{2}\cdots\mu_{p}}
\equiv\Gamma_{[\mu_{1}}\Gamma_{\mu_{2}}\cdots\Gamma_{\mu_{p}]}
\end{equation}
where
\begin{equation}
\Gamma_{\mu}=\rho(e_{\mu})\in\mbb R(16)
\label{gammaRep}
\end{equation}
are known as ``$\gamma$-matrices'', and the square bracket indicates
anti-symmetrization of the indices, with a prefactor $1/p!$. We shall use the
notation $\Gamma_{\mu_{1}\cdots\mu_{p}}=I_{16\times 16}$ for $p=0$.

The Clifford multiplication ``$\cdot$'' between $u=\sum_{\mu}C^{\mu}e_{\mu}
\in\wedge^{1}\mbb R^{8}\subset Cl(8)$ and any element $w\in Cl(8)\cong
\wedge^{*}\mbb R^{8}$ can be identified with an operation on the wedge algebra:
\begin{equation}
u\cdot w\longleftrightarrow u\wedge w - i_{u}(w),
\label{multiplication}
\end{equation}
where the interior product $i_{u}(w)$ is defined by the linear map $i_{u}:
\wedge^{p}\mbb R^{8}\rightarrow \wedge^{p-1}\mbb R^{8}$, via
\begin{equation}
i_{u}(u_{1}\wedge u_{2}\wedge\cdots\wedge u_{p})=\sum_{i=1}^{p}
(-1)^{i+1}\langle u_{i},u\rangle u_{1}\wedge\cdots\wedge\widehat{u_{i}}\wedge
\cdots\wedge u_{p}.
\label{contraction}
\end{equation}
Applying this to the matrix representation $\rho$, (\ref{multiplication})
becomes an identity between $\gamma$-matrices 
\begin{equation}
\Gamma_{\mu}\Gamma_{\nu_{1}\nu_{2}\cdots\nu_{p}}=\Gamma_{\mu\nu_{1}\nu_{2}
\cdots\nu_{p}}-g_{\mu\nu_{1}}\Gamma_{\nu_{2}\nu_{3}\cdots\nu_{p}}+
g_{\mu\nu_{2}}\Gamma_{\nu_{1}\nu_{3}\cdots\nu_{p}}+\cdots+(-1)^{p}
g_{\mu\nu_{p}}\Gamma_{\nu_{1}\cdots\nu_{p-1}}.
\label{gammaID}
\end{equation}

Sometimes we need to fix a particular basis and construct the
$\gamma$-matrices $\Gamma_{\mu}$ explicitly. Our convention of choosing such
matrices is as follows. Since $\mbb R(16)\cong\mbb R(2)\otimes\mbb R(2)
\otimes \mbb R(2)\otimes \mbb R(2)$, $\Gamma_{\mu}$ can be expressed by a
4-fold tensor product of some basis in $\mbb R(2)$. Thus, we take a basis of
$\mbb R(2)$ to be
\begin{equation}
I=\left (
\begin{array}{cc}
1 & 0 \\
0 & 1
\end{array}
\right ),\quad
\sigma_{1}=\left (
\begin{array}{cc}
0 & 1 \\
1 & 0
\end{array}
\right ),\quad
\sigma_{3}=\left (
\begin{array}{cc}
1 & 0 \\
0 & -1
\end{array}
\right ),\quad
\epsilon=\left (
\begin{array}{cc}
0 & 1 \\
-1 & 0
\end{array}
\right ).
\end{equation}
It is easy to checks that the $16\times 16$ matrices
\begin{equation}
\begin{array}{ll}
\Gamma_{1}=\epsilon\otimes\epsilon\otimes\epsilon\otimes\sigma_{1},\quad\quad &
\Gamma_{2}=I\otimes\sigma_{1}\otimes\epsilon\otimes\sigma_{1}, \\
\Gamma_{3}=I\otimes\sigma_{3}\otimes\epsilon\otimes\sigma_{1}, &
\Gamma_{4}=\sigma_{1}\otimes\epsilon\otimes I\otimes\sigma_{1}, \\
\Gamma_{5}=\sigma_{3}\otimes\epsilon\otimes I\otimes\sigma_{1}, &
\Gamma_{6}=\epsilon\otimes I\otimes\sigma_{1}\otimes\sigma_{1}, \\
\Gamma_{7}=\epsilon\otimes I\otimes\sigma_{3}\otimes\sigma_{1}, &
\Gamma_{8}=I\otimes I\otimes I\otimes\epsilon
\end{array}
\label{gamma}
\end{equation}
obey the relations $\{\Gamma_{\mu},\,\Gamma_{\nu}\}=-2\delta_{\mu\nu}$.

In our convention (\ref{gamma}), the matrices $\Gamma_{\mu}$ are all
anti-symmetric. More generally we have
\begin{equation}
(\Gamma_{\mu_{1}\mu_{2}\cdots\mu_{p}})^{T}=(-1)^{p}\,\Gamma_{\mu_{p}\cdots
\mu_{2}\mu_{1}}=(-1)^{\frac{p(p+1)}{2}}\Gamma_{\mu_{1}\mu_{2}\cdots\mu_{p}}.
\label{gammaSymm}
\end{equation}
Thus $\Gamma_{\mu_{1}\mu_{2}\cdots\mu_{p}}$ is anti-symmetric when $p\equiv 1,
2$ (mod 4) and symmetric when $p\equiv 3,4$ (mod 4). Consequently, the matrix
representation of the ``volume element'' $\omega\equiv e_{1}\wedge e_{2}
\wedge\cdots\wedge e_{8}\in Cl(8)$, namely
\begin{equation}
\rho(\omega)=\Gamma_{1}\Gamma_{2}\cdots\Gamma_{8}\equiv\Gamma_{9},
\end{equation}
has the symmetric property $(\Gamma_{9})^{T}=\Gamma_{9}$. In fact $\Gamma_{9}$
is diagonal in our basis:
\begin{equation}
\Gamma_{9}=I\otimes I\otimes I\otimes\sigma_{3}=\left(
\begin{array}{cc}
I_{8\times 8} & 0 \\
0 & -I_{8\times 8}
\end{array}
\right).
\label{gamma5}
\end{equation}

(\ref{gamma5}) shows that the irreducible $Cl(8)$ module $S\cong\mbb
R^{16}$ has a decomposition $S=S^{+}\oplus S^{-}$ into the $\pm1$ eigenspaces
$S^{\pm}$ of $\Gamma_{9}$. Of course neither of these eight-dimensional
eigenspaces are invariant under the action of $Cl(8)$. To be a little more
explicit, notice that $\Gamma_{9}\Gamma_{\mu_{1}\cdots\mu_{p}}
=(-1)^{p}\Gamma_{\mu_{1}\cdots\mu_{p}}\Gamma_{9}$, we have
\begin{equation}
\Gamma_{\mu_{1}\cdots\mu_{p}}\sim\left\{
\begin{array}{ll}
\left(
\begin{array}{cc}
\star & 0\\
0 & \star
\end{array}
\right),\quad & p={\rm even}\vspace{.3cm} \\
\left(
\begin{array}{cc}
0 & \star \\
\star & 0
\end{array}
\right),\quad & p={\rm odd}
\end{array}
\right.\quad {\rm on}\quad S=\left(
\begin{array}{c}
S^{+} \\
S^{-}
\end{array}
\right),
\label{block}
\end{equation}
so for odd $p$ the matrix $\Gamma_{\mu_{1}\cdots\mu_{p}}$ swaps $S^{\pm}$.
Nevertheless, if one considers a subalgebra of $Cl(8)$ spanned by some even
elements $a=e_{\mu_{1}}\wedge\cdots\wedge e_{\mu_{2k}}$, then the matrix
representation $\rho(a)=\Gamma_{\mu_{1}\cdots\mu_{2k}}$ will keep both the
subspaces $S^{\pm}\subset S$ invariant.

At this point we consider a linear space $\wedge^{2}\mbb R^{8}$
spanned by elements of the form
$L_{\mu\nu}=\frac{1}{2}e_{\mu}\wedge e_{\nu}=\frac{1}{4}(e_{\mu}e_{\nu}-
e_{\nu}e_{\mu})$. This spcae forms a Lie algebra under the
bracket
\begin{equation}
[L_{\mu\nu},\,L_{\alpha\beta}]\equiv
L_{\mu\nu}\cdot L_{\alpha\beta}-L_{\alpha\beta}\cdot L_{\mu\nu},
\end{equation}
where ``$\cdot$'' again stands for the Clifford multiplication. In fact, a
simple computation shows that
\begin{equation}
[L_{\mu\nu},\,L_{\alpha\beta}]=g_{\mu\alpha}L_{\nu\beta}+g_{\nu\beta}
L_{\mu\alpha}-g_{\nu\alpha}L_{\mu\beta}-g_{\mu\beta}L_{\nu\alpha},
\end{equation}
so $\{L_{\mu\nu}\}$ generates the Lie algebra of
$Spin(8)$. According to the previous discussion, $S^{\pm}$
can be considered as $Spin(8)$-modules, and actually they are two
inequivenlent irreducible modules of $Spin(8)$.
The representation $\rho(L_{\mu\nu})=\frac{1}{2}\Gamma_{\mu\nu}$
of (the Lie algebra of) $Spin(8)$ then decomposes into two irreducible ones:
$\rho=\rho^{+}\oplus\rho^{-}$. $\rho^{+}$ is the spin representation with
positive chirality and $\rho^{-}$ the spin representation with negative
chirality. That $\rho^{\pm}$ are inequivalent stems from the ``central
element'' $\Gamma_{9}=\rho(\omega)=\pm 1$ having different values on
$S^{\pm}$.

So far we have only constructed two irreducible spin representations
of $Spin(8)$, $\rho^{\pm}$, acting on $S^{+}={\bf 8}_{s}$ and
$S^{-}={\bf 8}_{c}$, respectively. There is another inequivalent
 eight-dimentional irreducible representation of $Spin(8)$, the so-called
  ``vector representation'' $\rho_{v}$, which will act on the vector space
 ${\bf 8}_{v}\equiv Span\{e_{\mu}\}\cong \wedge^{1}\mbb R^{8}$ adjointly:
\begin{equation}
\rho_{v}(L_{\mu\nu})(e_{\alpha})\equiv [L_{\mu\nu},e_{\alpha}].
\label{Adj}
\end{equation}
The matrix elements of $\rho_{v}(L_{\mu\nu})$ are determined by the following
commutative relations:
\begin{equation}
[L_{\mu\nu},e_{\alpha}]=g_{\mu\alpha}e_{\nu}-g_{\nu\alpha}e_{\mu}.
\label{vector}
\end{equation}
It follows that the image of $Spin(8)$ under $\rho_{v}$ is isomorphic to
$SO(8)$. The existence of the three inequivalent irreducible 8-dimensional
modules ${\bf 8}_{s}$, ${\bf 8}_{c}$ and ${\bf 8}_{v}$ is often summarized
as the triality of the $Spin(8)$-representations.

As a natural extension of the vector representation $\rho_{v}$, it is 
possible to construct tensor representations of $Spin(8)$ on $\wedge^{p}
\mbb R^{8}$. One verifies by induction that (\ref{vector}) is extended to
\begin{equation}
[L_{\mu\nu},e_{\alpha_{1}}\wedge\cdots\wedge e_{\alpha_{p}}]=
M^{\beta_{1}\cdots\beta_{p}}_{\alpha_{1}\cdots\alpha_{p}}(L_{\mu\nu})
e_{\beta_{1}}\wedge\cdots\wedge e_{\beta_{p}}
\label{tensor}
\end{equation}
with some adjoint matrices $M(L_{\mu\nu})$. This therefore defines a tensor
representation $\wedge^{p}\rho_{v}$ of $Spin(8)$. Alternatively,
$\wedge^{p}\rho_{v}$ may also be obtained by considering tensor product
of the spin representations $\rho^{\pm}$.

To see this, we first need to establish an isomorphism between the spaces
$(S^{+}\oplus S^{-})\otimes (S^{+}\oplus S^{-})$ and $\wedge^{*}\mbb R^{8}$.
Note that both spaces have the same dimensions: $16\times 16=2^{8}$. By
choosing an orthogonal basis $\{v_{A}\}_{1\leq A\leq 16}$ of  $S=S^{+}\oplus
S^{-}$, we associate each $v_{A}\otimes v_{B}\in S\otimes S$ to an element
of $\wedge^{*}\mbb R^{8}$ as follows:
\begin{equation}
v_{A}\otimes v_{B}\longleftrightarrow\bigoplus_{p=0}^{8}\,\langle
v_{A},\Gamma^{\mu_{1}\cdots\mu_{p}}v_{B}\rangle\, e_{\mu_{1}}\wedge
\cdots\wedge e_{\mu_{p}}.
\label{isom}
\end{equation}
This correspondence is 1:1 since if $v_{A}\otimes v_{B}$, $v_{C}\otimes
v_{D}$ are associated to the same element of $\wedge^{*}\mbb R^{8}$, then
we must have $\langle v_{A}, \Gamma^{\mu_{1}\cdots\mu_{p}}v_{B}\rangle\equiv
\langle v_{C}, \Gamma^{\mu_{1}\cdots\mu_{p}}v_{D}\rangle$ for all $p$ and,
by irreducibility of the Clifford
group\footnote{Clifford group $G_{d}\subset Cl(d)$ in $d$-dimensions is a
finite group whose generators can be presented by the abstract elements
$\{e_{1},\cdots, e_{d},-1\}$ subject to the relation that $-1$ is central
and that $(-1)^{2}=1,e_{i}^{2}=-1$ and $e_{i}e_{j}=(-1)e_{j}e_{i}$ for
all $i\neq j$.} acting on $S$, the two
matrix elements must be orthogonal and never identical to each
other, unless $v_{A}=v_{C}$, $v_{B}=v_{D}$.

Now we can use the isomorphism $S\otimes S\cong \wedge^{*}\mbb R^{8}$
specified by (\ref{isom}). We see that the action of any group
element $g\in Spin(8)$ on $v_{A}$, {\it i.e.} $v_{A}\rightarrow
\tilde{v}_{A}\equiv(\rho^{+}\oplus\rho^{-})(g)\circ v_{A}$, will induce
two equivalent actions on $v_{A}\otimes v_{B}$ and on its image in
$\wedge^{*}\mbb R^{8}$. The first action is simply
$(\rho^{+}\oplus\rho^{-})\otimes(\rho^{+}\oplus\rho^{-})(g):
v_{A}\otimes v_{B}\rightarrow \tilde{v}_{A}\otimes\tilde{v}_{B}$.
The second action, when restricted to the components $T^{\mu_{1}\cdots
\mu_{p}}\equiv\langle v_{A},\Gamma^{\mu_{1}\cdots\mu_{p}}v_{B}\rangle
\in\wedge^{p}\mbb R^{8}$, is determined by $T^{\mu_{1}\cdots
\mu_{p}}\rightarrow\tilde{T}^{\mu_{1}\cdots
\mu_{p}}\equiv\langle\tilde{v}_{A},\Gamma^{\mu_{1}\cdots\mu_{p}}
\tilde{v}_{B}\rangle$, which in turn gives rise to the adjoint action
$\Gamma^{\mu_{1}\cdots\mu_{p}}\rightarrow\rho(g)^{T}\Gamma^{\mu_{1}\cdots
\mu_{p}}\rho(g)=\rho(g)^{-1}\Gamma^{\mu_{1}\cdots\mu_{p}}\rho(g)$, leading
to the tensor representation $\wedge^{p}\rho_{v}$. It follows that
\begin{equation}
(\rho^{+}\oplus\rho^{-})\otimes(\rho^{+}\oplus\rho^{-})\cong
\bigoplus_{p=0}^{8}{\wedge}^{p}\rho_{v}\cong 2\left(1\oplus
\rho_{v}\oplus{\wedge}^{2}\rho_{v}\oplus{\wedge}^{3}\rho_{v}\right)
\oplus {\wedge}^{4}\rho_{v}.
\label{isomR}
\end{equation}

The isomorphism discussed above gives an identity known as the Fierz
rearrangement formula. The vectors $v_{A}$, $v_{B}$ in (\ref{isom})
can be replaced by arbitrary spinors $\phi=\phi^{A}v_{A},\psi=\psi^{A}v_{A}
\in S$. On the left hand side of this correspondence, we have the tensor
product $\phi\otimes\psi$ with components $\phi^{A}\psi^{B}$, which can be
viewed as a $16\times 16$ matrix acting on $S$. The right hand side can also
be considered as such a matrix if we replace the Clifford elements
$e_{\mu_{1}}\wedge\cdots\wedge e_{\mu_{p}}$ by their $\gamma$-matrix
representation $\Gamma_{\mu_{1}\cdots\mu_{p}}$. Since these two matrices are
the same object, we must have
\begin{equation}
\phi^{A}\psi^{B}=\frac{1}{16}\sum_{p=0}^{8}\frac{1}{p!}
(\phi^{T}\Gamma^{\mu_{1}\cdots
\mu_{p}}\psi)\,{\Gamma_{\mu_{1}\cdots\mu_{p}}}^{AB}
\label{fierz}
\end{equation}
here $(\Gamma_{\mu_{1}\cdots\mu_{p}})^{AB}$ denotes the matrix-element of
$\Gamma_{\mu_{1}\cdots\mu_{p}}$ in the basis $v_{A}$. The coefficients
$\frac{1}{p!}$ in (\ref{fierz}) are introduced so as to ensure that the sum
runs over each of the basis elements of $\wedge^{*}\mbb R^{8}$ exactly once,
and the factor $\frac{1}{16}$ comes from a group-theoretical consideration,      
which is nothing but the inverse of the dimension of the irreducible
representation for the Clifford group. That this factor must be equal to
$\frac{1}{16}$ may also be checked by taking the trace of (\ref{fierz}): From
(\ref{block}) we recall that for odd $p$, the matrix $\Gamma_{\mu_{1}\cdots
\mu_{p}}$ is always off-diagonal, thus having a vanishing trace. For even
$p>0$, the cyclic property of the trace ${\rm Tr}(\Gamma_{
\mu_{1}\mu_{2}\cdots\mu_{p}})={\rm Tr}(\Gamma_{\mu_{2}\cdots\mu_{p}\mu_{1}})$
together with the $\gamma$-matrix identity $\Gamma_{\mu_{1}\mu_{2}\cdots
\mu_{p}}=-\Gamma_{\mu_{2}\cdots\mu_{p}\mu_{1}}$ also gives ${\rm Tr}
(\Gamma_{\mu_{1}\mu_{2}\cdots\mu_{p}})=0$. So when taking the trace,
only the first term ($p=0$) in the right hand side of (\ref{fierz}) survives
and it takes the value $\frac{1}{16}\phi^{T}\cdot\psi{\rm Tr}(I_{16\times 16})
=\phi^{T}\cdot\psi$, which agrees exactly with the trace of the left hand
side. As an aside, note that this kind of argument allows us to write down a
trace formula for the $\gamma$-matrices: Multiplying (\ref{fierz}) by
${\Gamma^{\nu_{1}\cdots\nu_{q}}}_{AB}=(-1)^{[\frac{q+1}{2}]}{\Gamma^{\nu_{1}\cdots\nu_{q}}}_{BA}$, the
left hand side becomes $\phi^{T}\Gamma^{\nu_{1}\cdots\nu_{q}}\psi$, while the right
hand side is $\sum_{p=0}^{8}\frac{(-1)^{[\frac{q+1}{2}]}}{16p!}
(\phi^{T}\Gamma^{\mu_{1}\cdots\mu_{p}}\psi){\rm Tr}(\Gamma_{\mu_{1}\cdots
\mu_{p}}\Gamma^{\nu_{1}\cdots\nu_{q}})$, and this gives
\begin{equation}
{\rm Tr}(\Gamma_{\mu_{1}\cdots\mu_{p}}\Gamma^{\nu_{1}\cdots\nu_{q}})
=16 p!(-1)^{[\frac{p+1}{2}]}\delta_{pq}\delta^{\nu_{1}}_{[\mu_{1}}\cdots
\delta^{\nu_{p}}_{\mu_{p}]}.
\label{gammaTr}
\end{equation}
For example we have ${\rm Tr}(\Gamma_{\mu}\Gamma^{\nu})=-16\delta^{\nu}_{\mu}$,
${\rm Tr}(\Gamma_{\mu\nu}\Gamma^{\alpha\beta})=-16(\delta_{\mu}^{\alpha}
\delta_{\nu}^{\beta}-\delta_{\mu}^{\beta}\delta_{\nu}^{\alpha})$, {\it etc.}.

We end this appendix with a few remarks. If $\phi=\psi$ has a definite
chirality, then many terms in the sum (\ref{fierz}) will vanish. Such terms
correspond to $p\equiv 1,2$ (mod 4) when the $\gamma$-matrices are
anti-symmetric or $p=$
odd when the $\gamma$-matrices map $\phi$ into some spinors with opposite
chirality, which are orthogonal to $\phi^{T}$. In this case the Fierz
reaarangement formula gets much simplified:
\begin{equation}
\phi\phi^{T}=\frac{1}{16}\left((\phi^{T}\phi)I_{16\times 16}+(\phi^{T}
\Gamma_{9}\phi)\Gamma_{9}+
\frac{1}{4!}(\phi^{T}\Gamma^{\mu\nu\alpha\beta}\phi)\Gamma_{\mu\nu\alpha
\beta}\right).
\label{fierz1}
\end{equation}
Thus, since $\Gamma_{9}\phi=\pm\phi$ for $\phi\in S^{\pm}$, we have
\begin{equation}
\phi\phi^{T}=\frac{1}{16}\left((\phi^{T}\phi)(I_{16\times 16}\pm
\Gamma_{9})+\frac{1}{4!}(\phi^{T}\Gamma^{\mu\nu\alpha\beta}\phi)
\Gamma_{\mu\nu\alpha\beta}\right),\quad\phi\in S^{\pm}.
\label{fierz1-1}
\end{equation}
Clearly (\ref{fierz1-1}) defines a projector from $S$ onto its one-dimensional
subspace spaned by $\phi$. Moreover, if $\phi\neq\psi$ but they still have
the same definite chirality -- say, both of them are in $S^{+}$ or in $S^{-}$,
then by symmetrizing (\ref{fierz}) we get
\begin{equation}
\phi^{A}\psi^{B}+\phi^{B}\psi^{A}=\frac{1}{8}\left((\phi^{T}\psi)
(\delta^{AB}\pm{\Gamma_{9}}^{AB})+\frac{1}{4!}(\phi^{T}\Gamma^{\mu\nu\alpha
\beta}\psi){\Gamma_{\mu\nu\alpha\beta}}^{AB}\right),
\label{fierz1-2}
\end{equation}
where ``$\pm$'' corresponds to $\phi$, $\psi\in S^{\pm}$, respectively.
We can also anti-symmetrize (\ref{fierz}) to derive, for $\phi$ and $\psi$
having the same chirality,
\begin{equation}
\phi^{A}\psi^{B}-\phi^{B}\psi^{A}=\frac{1}{8}\left(\frac{1}{2!}(\phi^{T}
\Gamma^{\mu\nu}\psi){\Gamma_{\mu\nu}}^{AB}
+\frac{1}{6!}(\phi^{T}\Gamma^{\mu\nu\alpha\beta\lambda\rho}\psi)
{\Gamma_{\mu\nu\alpha\beta\lambda\rho}}^{AB}\right).
\label{fierz1-3}
\end{equation}
The two terms in the right hand side of
(\ref{fierz1-3}) are in fact equal to each other upto a factor $\pm
\Gamma_{9}$ and we finally have $\phi\psi^{T}-\psi\phi^{T}=\frac{1}{16}
(\phi^{T}\Gamma^{\mu\nu}\psi)\Gamma_{\mu\nu}(1\pm\Gamma_{9})$, if both
$\phi$, $\psi$ are in $S^{\pm}$.

\end{document}